# Structure, composition and location of organic matter in the enstatite chondrite Sahara 97096 (EH3)

L. Piani[1], F. Robert[1], O. Beyssac[2], L. Binet[3], M. Bourot-Denise[1], S. Derenne[4], C. Le Guillou[5], Y. Marrocchi[6], S. Mostefaoui[1], J.N. Rouzaud[5], A. Thomen[1]

[1]Laboratoire de Minéralogie et Cosmochimie du Muséum, CNRS-MNHN, UMR7202, Case 52, 57 Rue Cuvier, 75005 Paris, France (contact : lpiani@mnhn.fr),
[2]Institut de minéralogie et de physique des milieux condensés, UMR 7590, CNRS-UPMC, Campus Jussieu, 4 place Jussieu - 75005 PARIS, France.
[3]Chimie ParisTech, Laboratoire de Chimie de la Matière Condensée de Paris, CNRS, UMR 7574, ENSCP, 11 rue Pierre et Marie Curie 75231 Paris cedex 05
[4]BioEMCo, CNRS-UPMC, UMR 7618, 4 place Jussieu, 75252 Paris, France.
[5]Laboratoire de Géologie de l'Ecole normale supérieure, CNRS-ENS, UMR 8538, 24 Rue Lhomond, 75231 Paris, France.
[6]Centre de Recherches Pétrographiques et Géochimiques - Nancy Université, CNRS, UPR 2300, 15 Rue Notre-Dame des Pauvres, BP20, 54501 Vandœuvre-lès-Nancy, France.

**ABSTRACT**

The insoluble organic matter (IOM) of an unequilibrated enstatite chondrite Sahara97096 has been investigated using a battery of analytical techniques. As the enstatite chondrites are thought to have formed in a reduced environment at higher temperatures than carbonaceous chondrites, they constitute an interesting comparative material to test the heterogeneities of the IOM in the Solar System and to constrain the processes that could affect IOM during Solar System evolution.

The Sahara97096 IOM is found *in situ*: as sub-micrometer grains in the network of fine-grained matrix occurring mostly around chondrules and as inclusions in metallic nodules, where the carbonaceous matter appears to be more graphitized. IOM in these two settings has very similar $\delta^{15}N$ and $\delta^{13}C$; this supports the idea that graphitized inclusions in metal could be formed by metal catalytic graphitization of matrix IOM.

A detailed comparison between the IOM extracted from a fresh part and a terrestrially weathered part of Sahara97096 shows the similarity between both IOM samples in spite of the high degree of mineral alteration in the latter.

The isolated IOM exhibits a heterogeneous polyaromatic macromolecular structure, sometimes highly graphitized, without any detectable free radicals and deuterium-heterogeneity and having mean H and N isotopic compositions in the range of values observed for carbonaceous chondrites. It contains some sub-micrometer-sized areas highly enriched in $^{15}N$ ($\delta^{15}N$ up to 1600‰).

These observations reinforce the idea that the IOM found in carbonaceous chondrites is a common component widespread in the Solar System. Most of the features of Sahara97096 IOM could be explained by the thermal modification of this main component.

# 1. INTRODUCTION

Insoluble organic matter (IOM) constitutes the major part of the carbon found in various primitive objects of the solar system such as carbonaceous chondrites or interplanetary dust particles (IDP). Acid treatments (HF-HCl) are commonly used to isolate the IOM from meteorites by dissolving most of the minerals. The chemical, macromolecular and isotopic compositions of the IOM have yielded new constraints on the origin and formation mechanisms of this IOM (*e.g.*, Robert and Epstein 1982; Kerridge et al. 1987; Alexander et al. 1998, 2007; Binet et al. 2004a, 2004b; Cody et al. 2008; Remusat et al. 2005, 2006, 2010; Bonal et al. 2006, 2007; Busemann et al. 2006, 2007; Gourier et al. 2008; Quirico et al. 2009; Derenne and Robert 2010, Duprat et al. 2010). However, some important issues are still debated in the literature, namely: (1) did the IOM originate from the circum-solar (*e.g.*, Remusat et al. 2007) or from the interstellar medium (*e.g.*, Busemann et al. 2006)? (2) Does a common precursor exist for the IOM present in a wide variety of primitive materials (*e.g.*, Quirico et al. 2009; Alexander et al. 2007)? (3) What is the influence of parent body processes on the evolution of the IOM (*e.g.*, Hanon et al. 1998; Yabuta et al. 2007; Quirico et al. 2009; Alexander et al. 2010; Remusat et al. 2010)?

It seems likely that some answers to these open issues may be found in looking at new, non-carbonaceous chondrite groups having distinct formation and evolutionary contexts. Since enstatite chondrites were formed in environments quite distinct from those of the carbonaceous chondrites (in terms of temperature and oxygen fugacity), the aim of the present study is to determine the macromolecular structure, the chemical and isotopic compositions of their IOM. The comparison with IOM from carbonaceous chondrites may bring new information on the nature and distribution of organic carbon in the solar system at the time of parent body accretion.

In previously published papers, the IOM from enstatite chondrites was compared to the IOM from carbonaceous chondrites on the basis of their chemical and/or isotopic compositions (Yang and Epstein 1983; Deines and Wickman 1985; Busemann et al. 2007; Alexander et al. 1998, 2007; Cody et al. 2008). In the present paper, the organic matter isolated from one of the most primitive enstatite chondrites (Weisberg and Prinz 1998), Sahara 97096 (hereafter SAH97096), is investigated by a battery of analytical techniques. This organic matter is considered to be the least subjected to post-accretion parent body metamorphism in enstatite chondrites, as petrologic type 3.1-3.4 was recently assigned to SAH97096 by comparing the signal from its IOM with Raman spectroscopy to IOM from different EH3 and EL3 (Quirico et al. 2011).

Different techniques were used to study whole rock polished sections and isolated IOM: (1) *in situ* observations of the polished sections were done by reflected light optical microscopy, scanning electron microscopy (SEM) and electron probe micro-analysis (EPMA), (2) hydrogen, carbon and nitrogen elemental composition of the IOM was measured by thermal conductivity - gas chromatography, (3) macromolecular properties of the IOM were documented by Curie-point pyrolysis coupled with gas chromatography-mass spectrometry (GC-MS) analysis and electron paramagnetic resonance (EPR), (4) structural organization of the IOM was documented by Raman microspectroscopy, (5) visual observation of the structural organization at the nanometer scale was achieved by high resolution transmission electron microscopy (HRTEM) and (6) isotopic and elemental distributions were imaged by secondary ion mass spectrometry (SIMS).

Some specific questions related to enstatite chondrites were investigated in this study, namely: *(i)* Is there evidence for any textural relationships between the organic matter and surrounding minerals? *(ii)* What are the structural, macromolecular and isotopic similarities and differences between the IOM from enstatite and carbonaceous chondrites? *(iii)* Is the enstatite chondrite IOM chemically and isotopically homogeneous? *(iv)* Are any heterogeneities in the IOM related to the petrographic location of the carbonaceous matter in the rock?

We will show in this paper that the IOM from the enstatite chondrite SAH97096 is comparable in many respects to the IOM isolated from the thermally processed carbonaceous chondrites. It is located in the fine grained matrix and as inclusions within metal nodules. Although the structural organisation is different, no isotopic differences were observed between the organic matter present in matrix and in metal. The IOM also contains small areas ($\leq 300$ nm$^2$) significantly enriched in $^{15}$N. Thus it seems that IOM was a common compound in the inner solar system, *i.e.* where chondrites formed. The observed variations in the molecular structures or in the chemical compositions of the disordered IOM isolated from various classes of meteorites can be simply accounted for by variations in their accretion or metamorphism temperatures.

## 2. EXPERIMENTAL

This study focused on the organic matter of SAH97096, an enstatite chondrite found in 1997 in the Sahara (country unknown). The meteorite (2.52 kg) and its pairing group weighing ~ 28 kg were classified as EH3 on the basis of their petrography (Weisberg and Prinz 1998). Two fragments of the meteorite – one from the outer part exposed to the terrestrial weathering and one from the fresh inner zone of the rock– were used to provide information on the effects of the terrestrial alteration on the IOM structures and compositions. The weathered part (60 g) is a rusty powder while the pristine part (17 g) from the inner zone is a solid rock with no visible alteration at the naked eye. Both fragments and polished sections were provided by the MNHN Paris. Sample preparation and analytical procedures are described below and summarized in the flow chart reported in Fig. 1.

### 2.1. Whole rock preparation and *in situ* analyses

*2.1.1. Sample preparation*

The fresh part of SAH97096 was used to perform *in situ* observations and analyses. Polished sections from the MNHN collection were used for optical observations and Raman spectroscopy measurements. In order to avoid the problem of hydrogen and nitrogen contamination introduced during polishing due to epoxy and lubricant, we prepared a "special" rock section for isotopic measurements. This section was cut and polished under alcohol with diamond spray (but without using epoxy) as follows : (1) a slice of the pristine part of SAH97096 (around one square centimeter with a thickness of $\approx$ 3 millimeter) was glued with Canada balsam over a mounting (Fig. 2 a) to be polished successively with 3, 1 and 0.25 μm diamond spray using exclusively alcohol instead of a common lubricant, (2) the polished slice was unstuck by heating and fixed with spackle to an aluminum support as shown on the Fig. 2 b, (3) the conductive connection between the sample and the aluminum support was provided by a link made with silver paint (Fig. 2 c), (4) the assemblage was finally coated with 20 nm of gold.

After NanoSIMS analyses and before SEM, Raman microscopy and EPMA, the section was re-polished with the 1 and 0.3 μm Al$_2$O$_3$ polishing powder to remove the possible diamond grains trapped in the porosity of the section and some metal oxidation, which might be created by the fine polishing under alcohol. The obtained polished surface is rougher and large holes appear within the chondrules.

*2.1.2. Distribution and structure of the organic matter in SAH97096*

The polished sections of SAH97096 were first documented by optical microscopy. SEM was used to produce backscattered electron (BSE) images –with a 20 kV accelerating voltage, a 10 nA electronic current and a working distance of ~8 or ~39 mm. Energy-dispersive X-ray spectrometry (EDS) was used to map the elemental composition of the different areas of the samples to provide indications on their chemical composition with a spatial resolution of ~1 µm.

The sub-micrometric grain size and/or low concentration of this organic matter within the mineral matrix preclude its detection by conventional techniques. An alternative is Raman microspectroscopy which is extremely sensitive to aromatic carbon phases, allowing the mapping of the organic matter within the matrix at a micrometric scale. Raman microspectroscopy mapping, either point by point or in dynamic line-scanning mode, was performed on different polished sections of SAH97096 (including the epoxy-free section) using a Renishaw inVia Raman Microspectrometer (IMPMC Paris). All analyses were obtained with an argon laser (514.5 nm; Laser Physics) focused through a Leica DMLM microscope with a 50x or 100x objective (numerical aperture = 0.75 or 0.85 respectively) yielding a lateral spatial resolution of ~ 1 µm with a ~1 mW laser power at the surface of the sample; for technical details regarding Raman mapping see Bernard et al. (2008). After map acquisition, each spectrum was compared with a reference spectrum to build a false-colour map using a direct classical least squares component (DCLS) analysis method provided by the Renishaw software "Wire 3.2". Long time acquisition spectra measured *in situ* for enstatite, graphite and in isolated IOM for disordered organic matter were used as reference spectra. A correlation index was calculated indicating the degree of similitude between complete similitude (correlation index = 1) and the total absence of similitude (correlation index = 0). If the correlation index with a given reference spectrum was higher than 0.7, the color representing the phase was given to the pixel. The majority of the maps were done in the range of wavelength going from 900 to 2900 $cm^{-1}$: only the organic matter bands (at about 1350 and 1600 $cm^{-1}$) and the major enstatite signal band (at ~ 1100 $cm^{-1}$) were recorded. The uncolored pixels correspond to phases that are poorly Raman active or that have no Raman bands in the spectral window used for mapping – mainly metal and sulfides. Most of these phases were identified with other techniques (SEM or optical microscopy). Moreover, possible effects of polishing on the structure of the disordered organic matter in the matrix were investigated by measuring some spectra of pristine organic matter located beneath transparent enstatite minerals following Beyssac et al. (2003b). No structural difference was observed between spectra from organic matter located beneath the enstatite or at the surface of the polished section, confirming the lack of polishing effects on the structure of such very disordered carbonaceous material.

The abundance of carbon dissolved or present as sub-micrometer inclusions within the Fe-Ni phases was measured with a CAMECA SX-100 electron microprobe (UPMC Paris). Two electron beam conditions were used: an accelerating potential of 15 kV and a current of 10 nA to analyze Co, Ni, Si, Fe and accelerating potential of 10 kV and a current of 300 nA to analyze Cr, P, S and C. Three NIST reference materials were used as standards of metal: SRM 665, 661 and 664 containing respectively 0.008, 0.39 and 0.87 weight percent of carbon. Again these samples were polished without epoxy, mounted over an aluminium support and gold-coated. In these conditions, the possible sources of carbon contamination are minimized for the reference materials as they were for the meteorite polished section. The results for the standards are given in Table 1 and discussed in the results section of the article (cf. §3.5.).

*2.1.3. Isotopic analysis of the organic matter in SAH97096 whole rock*

Nitrogen and carbon isotopic compositions and C, H, N and Si distributions were imaged and quantified by NanoSIMS 50 analyses on the polished section (without resin) previously studied by optical microscope. The NanoSIMS analytical conditions are similar to the ones of the second IOM session (see below in section 2.2.3), but performed on a 15 x 15 μm$^2$ surface divided into 128 x 128 pixels with a 6 pA primary ion beam (~250 nm) after 10 minutes of presputtering at 12 pA over a surface of 18 x 18 μm$^2$. This procedure, along with a relatively low primary current, permits to obtain a reasonable counting statistic for $^{12}C^{15}N^-$ for the total image *i.e.* with a statistic error of ≤ 2 ‰ for the $^{15}N/^{14}N$ ratio. The number of cycles depends on the images and varied from 100 to 350.

## 2.2. The Insoluble Organic Matter

*2.2.1. Isolation of the IOM*

After extraction of soluble compounds in water and organic solvents, the IOM was isolated from SAH97096 using the standard HF/HCl treatment (Gardinier et al. 2000). IOM isolated from the weathered sampled was subjected to $HBO_3$/HCl (0.9N/6N, 2:1 ratio by volume, 3hr at room temperature) in order to dissolve fluorides formed during the acid treatment. Thereafter, the IOM isolated from the preserved and altered parts of SAH97096 are referred to as IOMp and IOMa, respectively (subscripts p and a stand for *preserved* and *altered*).

*2.2.2. Bulk and molecular analysis of the IOM*

The elemental concentrations were determined by gas chromatography with a thermal conductivity detector (for H, C, N) and by ionometry for F at the SGS MULTILAB laboratory and the Analysis Central Centre of the CNRS (France).

The IOM was pyrolyzed through 650°C Curie-point pyrolysis by a Fisher 0316 M pyrolyser coupled with a GC-MS. The GC-MS used was a Hewlett Packard HP 5890 N gas chromatograph and Hewlett Packard HP 5973 mass spectrometer with electron energy 70 eV, ion source temperature 120°C, scanning from 40 to 650 a.m.u., 2 scan/s. Helium was used as carrier gas. The temperature of the GC oven was raised from 50°C to 300°C at 4°C.min$^{-1}$. A fused silica capillary column RTX-5sil MS (30 m x 0.25 mm internal diameter x 0.5 μm film) was used to separate the products of pyrolysis. Compounds were identified by GC-MS based on mass spectra, GC retention times, and comparison with library mass spectra.

The EPR measurements were carried out with a Bruker Elexsys E500 spectrometer (Chimie ParisTech, Paris) operating at a microwave frequency ~9.4GHz. The measurements were performed at room temperature and atmospheric pressure conditions. EPR allows the identification of species having unpaired electron spins like *e.g.* organic radicals or ferromagnetic minerals.

Point analyses on the IOM by Raman microspectroscopy were performed using the configuration described in § 2.1.2.

Transmission Electron Microscopy (TEM) images were acquired on a JEOL 2011 operating at 200kV and using a $LaB_6$ filament (UPMC Paris). The IOM powder was dispersed in ethanol on TEM grids covered by a lacey amorphous carbon films. HRTEM images were taken from the thinnest edges of the particles, across holes of the carbon film. The structural organization was directly imaged by using the "high resolution" mode (300-600 k$\times$ magnification, resolution of about 0.144 nm). In these images, each fringe represents the edge-on profile of a polyaromatic layer.

*2.2.3. Isotopic analyses of the IOM*

The hydrogen and nitrogen isotopic composition and the hydrogen, carbon, nitrogen and silicon distributions of the IOM were imaged and quantified by scanning ion imaging at high mass resolution with the NanoSIMS 50 (MNHN Paris). We performed these analyses on aliquots from the pristine and weathered fractions of SAH97096 (IOMp and IOMa) as well as on two terrestrial kerogens (a type I from the Green River Shale formation and a type III coal from the Mahakam Delta) used as reference samples. All samples were pressed in gold foils. We had three different sessions of analyses with a 16 keV $Cs^+$ primary ion beam. Each session consists of a set of ion images recorded during the same period with the same analytical conditions. For the three sessions, the type III kerogen was used as standard. This standard was measured regularly (bracketing at least each two ion images of the samples). The type I kerogen was used only in the first session. Data processing of the NanoSIMS images was performed with the L'IMAGE software package (Larry Nittler, Carnegie Institution Washington).

The errors given for the isotopic ratios include the statistical ion counting error and the reproducibility on the type III kerogen standard. The statistical ion counting error corresponds to the square root of the number of counts for each analysis (Poisson error) and the reproducibility corresponds to the standard deviation of the measured isotopic ratios for the kerogen III over one session. All the error bars are given at $2\sigma$.

Session #1 was conducted on both IOMa and IOMp by coupling multi-collection to magnetic peak switching in three steps having three distinct magnetic fields. The D/H ratio, $^{15}N/^{14}N$ ratio and H-, C-, N-elemental distributions were measured in the same analytical session in order to observe possible relations between hydrogen and nitrogen isotopic anomalies, often observed in IOM from carbonaceous chondrites (Busemann et al., 2006; Thomen et al., 2009). The ions measured for each magnetic field were: $H^-$ and $D^-$ on the detectors 1 and 2 with the first magnetic field, $^{12}C^{14}N^-$ and $^{12}C^{15}N^-$ on detectors 4 and 5 with the second magnetic field and $^{13}C^-$ and $^{13}CH^-$ on the detectors 2 and 3 with the third magnetic field. In order to separate the major interferences ($^{13}C^-$ from $^{12}CH^-$ and $^{13}C^{14}N^-$ from $^{12}C^{15}N^-$), the mass resolution power of the machine defined by $MRP_{cameca} = 0.25 \cdot R \cdot \Delta M_{10-90}$, where R is the radius and $\Delta M_{10-90}$ is the mass span of 80% of the peak flank, see Fletcher et al. (2008) for details, was up to 8500. A 10-15 minutes pre-sputtering was done with a ~ 125 pA primary beam on a 25 x 25 $\mu m^2$ surface to eliminate superficial contamination and to approach the sputtering steady-state regime. Then, a primary ion beam of ~ 50 pA giving on a spatial resolution of ~ 500 nm was swept over a surface of 20 x 20 $\mu m^2$ divided into 256 x 256 pixels. The time of measurement per pixel in one cycle was 10 $\mu s$ and each image is constituted of 80 cycles to obtain a satisfying number of counts for $D^-$. Using these parameters, the time to obtain one ion image was 7 hours. For this session the reproducibility for the isotopic ratios in the terrestrial standard, the type III kerogen, is given in Table 2. Isotopic composition are reported as ratios and/or as delta values, $\delta D$, $\delta^{15}N$ and $\delta^{13}C$, representing the deviation from the terrestrial reference values in part per thousand. The terrestrial reference values are standard mean ocean water for hydrogen (D/H = 155.76 $\times 10^{-6}$), the marine carbonate standard Pee Dee Belemnite for carbon ($^{13}C/^{12}C$ = 1.2375 $\times 10^{-2}$) and atmospheric air for nitrogen ($^{15}N/^{14}N$ = 3.6765 $\times 10^{-3}$).

Session #2, where the IOM and the whole rock were analyzed (see the following section 2.2.3.), was performed with the aim of improving the precision on the nitrogen isotopic ratios and searching for possible relations between $^{15}$N-enrichments and elemental silicon. Ions collected by muticollection on the 5 electron multiplier detectors were $^{12}$C$^-$ (detector #1), $^{13}$C$^-$ or $^{12}$CH$^-$ depending on the analysis (detector #2), $^{12}$C$^{14}$N$^-$ (detector #3), $^{12}$C$^{15}$N$^-$ (detector #4) and $^{28}$Si (detector #5). For this session, the mass resolution power (MRP$_{cameca}$) for $^{12}$C$^{14}$N$^-$ was around 10,000. A 12 pA primary ion beam (~ 300 nm of diameter) was swept on 20 x 20 µm$^2$ surfaces and each surface was imaged 150 times (around 3 hours of measurement). Since the spatial homogeneity of the D/H distribution was observed during the first session, the H$^-$ and D$^-$ were not measured during this second session. The use of a single magnetic field drastically reduced the time of analyses for session #2 in comparison with the session #1. For this session #2, the reproducibility on the terrestrial standard (the type III kerogen) was ± 1.6 % for the $^{15}$N/$^{14}$N ratio (c.f. Table 2).

The data processing employed to quantify the H$^-$, C$^-$ and N$^-$ variations in the images for session #1 and #2 is described below. For each image, only the last 10 to 50 cycles were used to insure that the extraction steady state is obtained for all ion yields. The H/C and N/C ratios for each image were corrected for instrumental mass fractionation using the kerogen analyses. Then, the part of the image containing the carbonaceous matter was divided into regions of interest (ROI) having approximately the size of the primary beam. Finally, histograms of frequency were calculated containing the values of all the ROI of a same sample.

Session #3 was performed to test the validity of our observations of small $^{15}$N-rich anomalies in the IOM. Because the $^{12}$C$^{15}$N$^-$ used to measure the $^{15}$N/$^{14}$N ratio is under abundant, it is highly exposed to isobaric interferences. Ignoring the $^{13}$C$^{14}$N$^-$ interference that is clearly resolved at our mass resolving power, the major interference at 27 amu is the boron oxide ion $^{11}$B$^{16}$O$^-$; the mass resolution power (M/ΔM where ΔM is the mass difference between $^{11}$B$^{16}$O$^-$ and $^{12}$C$^{15}$N$^-$) required to separate $^{11}$B$^{16}$O$^-$ from $^{12}$C$^{15}$N$^-$ is theoretically 6600. Following the protocol proposed by Thomen et al. 2010, we measured $^{12}$C$^-$, $^{16}$O$^-$, $^{12}$C$^{14}$N$^-$, $^{12}$C$^{15}$N$^-$ and $^{11}$B$^{18}$O$^-$ distributions in the IOM with the 5.8 pA Cs$^+$ primary ion beam over 20 x 20 µm$^2$ surfaces and each surface was imaged 110 times with a mass resolution power (MRP$_{cameca}$) of 10,000-12,000. No correlation was found between the $^{15}$N/$^{14}$N ratio and the $^{11}$B$^{18}$O$^-$ rich areas. This observation excludes the possibility that any of the $^{15}$N anomaly could be caused by the boron oxide isobaric interference.

## 3. RESULTS

### 3.1. Organic matter occurrences within SAH97096

Using optical microscopy in reflected light and SEM-EDX imaging on the polished sections of SAH97096, we observed the main mineral assemblages reported from other unequilibrated enstatite chondrites (*e.g.*, Keil 1968). The meteorite consists of an assemblage of enstatite chondrules, chondritic fragments and opaque nodules with a fine grained matrix filling the interstices. Rubin et al. (2009) distinguished three distinct types of matrix in EH3 chondrites consisting of: (1) metal, sulfides and phosphides in coarse fragments and polycrystalline assemblages, (2) coarse-grained silicate assemblages and (3) a fine-grained component enriched in volatile elements with sub-micrometric grains. These distinct types of matrix are present in SAH97096. The fine-grained matrix is mainly located around the chondrules. As this article is focused on the IOM, we considered the fined-grained matrix as the most promising area for containing such IOM grains and it will be termed "matrix" in the following.

The main crystalline phases in SAH97096 are near pure enstatite, silicon-rich kamacite, nickel-poor troilite, feldspathic glass, SiO$_2$, Mg-olivine, other sulfides like niningerite (Fe,Mg,Mn)S, oldhamite CaS, daubreelite $FeCr_2S_4$, sphalerite (Fe,Zn)S, djerfisherite $(K,Na)_6(Fe,Ni,Cu)_{25}S_{26}Cl$, the silicide perryite $(Ni,Fe)_5(Si,P)_2$, and the phosphide schreibersite $(Fe,Ni)_3P$. The chondrules consist mostly of enstatite phenocrysts in a mesostasis of plagioclase-like composition. We observed poikilitically enclosed forsteritic olivine in some enstatite chondrules. Sulfides are either associated with kamacite in nodules or isolated in the matrix. Schreibersite is found as inclusions in the different types of sulfides, in the metal and isolated in the matrix. The perryite is mainly imbricated with troilite in association with metal. Nodules are often a complex association of troilite-perryite, kamacite, sulfides and various inclusions. Note that the petrographic description holds only for the pristine part of SAH97096, the corresponding textures of the altered part being barely recognizable.

SAH97096 exhibits localized shock features (molten metal and sulfide veins, melt pockets etc.). Quirico et al. (2011) proposed a S3-4 shock degree based on criteria defined for enstatite chondrites by Rubin and Scott (1997). For the following *in situ* observation and analysis, we chose areas with no apparent shock features.

As found in previous studies on enstatite chondrites (Weisberg and Prinz 1998; Rubin et al. 2009), we located the IOM at two main locations in SAH97096.

<u>Location 1</u> The organic matter was found as intimately mixed with silicates, sulfides and phosphides and located within the matrix areas, mainly occurring as thin rinds (10-30μm) around chondrules (Fig. 3). Raman microspectroscopy was used for the mapping of the organic matter within the matrix at a micrometric scale. Among the other phases contained in SAH97096, only enstatite has a well defined Raman signature. Opaque phases like sulfides, kamacite, schreibersite and perryite are weakly Raman active (cf. Fig. 4), but their identification was done easily with optical and scanning electron microscopes. In the Raman maps, the first order vibration bands (D and G bands) of the disordered carbonaceous matter are clearly observed all over the largest matrix areas. At the scale of the Raman analysis, IOM is widespread in the matrix and between chondrules, chondrule fragments and opaque nodules. Fig. 4 shows one of these Raman maps along with spectra of disordered IOM in matrix area, graphitized IOM inclusion within metal and some representative spectra of minerals. The spectra of minerals were recorded with higher laser intensity (x 3) and a longer acquisition time (x 5) than used during the map acquisition. Subtle variations in the width of the IOM Raman bands and more generally in the shape of the Raman spectra of IOM among all the spectra in the map could be observed, suggesting that minor structural variations of the IOM occur at the Raman scale of observation (μm). Additionally, the Raman spectra obtained for the IOM are similar to those measured in SAH97096 (Quirico et al., 2011) confirming the structural homogeneity at the Raman scale. Some rare occurrences of more organized IOM with spectra close to the spectra of graphitized IOM in metal are observed. All these occurrences seem to be spatially linked with metal or sulfide.

<u>Location 2</u> The other host for the organic matter is the metal and metal-sulfide nodules and was first mentioned in SAH97096 by Weisberg and Prinz (1998). Such occurrences were already studied in some other enstatite, ordinary and carbonaceous chondrites (Rubin and Keil 1983, Mostefaoui et al. 1998, 2000, 2005) and in primitive achondrites (Charon et al. 2009, 2010). In the St Marks enstatite chondrite, graphite was found showing book-like morphology (*i.e.* with well stacked sheets, like the pages of a closed book) in association with metal, silicates or troilite (Mostefaoui et al. 2005). Such book shaped morphology of graphite is found exclusively in silicates in the Abee enstatite chondrite breccia. In SAH97096, we did not find this book shaped graphite but spherulitic or granular graphite-like and rarer lamellar graphite-like inclusions within the kamacite or, to a lesser part, within the troilite-perryite assemblages. The largest occurrence of graphitized carbon found in metal (diameter ~ 25 μm) is shown in a back scattered electron (BSE) picture (Fig. 5a). Two other graphitized carbon inclusions are shown on secondary electron pictures (Fig. 5b and 5c). In the SEM pictures of the resin-free polished section, which correspond to a total area of ~ 6.5 mm$^2$, we observed 51 metallic nodules with a diameter ≥ 100 μm. The largest diameter reaches 500 μm. Eleven of these nodules, which present an assemblage of kamacite, troilite and perryite and schreibersite, contain visible graphitized inclusions in the kamacite. Two nodules possess inclusions within their troilite-perryite imbrications. Considering the mean size of the inclusions (1-2 microns) and the fact that only a cross section of each nodule is visible on the polished section, it is plausible that each nodule in SAH97096 contains such inclusions of graphitized material. Raman microspectroscopy on the graphite-like inclusions indicates that this material is highly graphitized, with an intense and very sharp G band and low intensity $D_1$ and $D_2$ bands (Fig. 4). All the spectra were analyzed at the surface of the polished section as the surrounding kamacite is highly opaque and does not allow for measuring graphite at depth within the thin section thickness. As shown in Mostefaoui et al. (2000) and Ammar et al. (2011), the polishing could induce structural defects to graphite inclusions and because the laser does not penetrate in the black carbon, Raman spectroscopy is highly sensitive to the surface defects. Therefore, the defect bands may be an artefact introduced by the polishing.

EPMA was used to obtain a semi-quantitative estimate of the concentration of carbon dissolved or present as submicrometer-inclusions in the metal of the nodules. The carbon concentration obtained for 22 different analyses on 16 different nodules were compared to the three NIST metal analyses (Table 1). Of the 16 nodules, only 7 have visible organic inclusions. Using the regression line obtained from the NIST references and taking the standard deviation of each reference into account, we calculated the C concentration for each analysis in the metal. With a carbon concentration of -0.08 ± 0.52 wt.%, the kamacite contains virtually no carbon. Only 4 of the 22 metal analyses give a carbon concentration higher than the lowest carbon concentration of the NIST references (0.08 wt.% of C). There is no correlation between the presence of visible organic inclusion and the C concentration of the kamacite.

### 3.2. Isotopic and elemental distributions of the *in situ* IOM imaged with the NanoSIMS

Isotopic and elemental NanoSIMS imaging was performed in similar types of locations – matrix and metallic nodules – where the IOM was detected on the resin-free polished section. The N and C isotopic values of the IOM for each image are given in Table 3. The 15 x 15 μm$^2$ NanoSIMS images of the matrix show the distribution of the IOM: it appears as isolated micro- to sub-micrometer sized grains enriched in C, H and N. NanoSIMS images of the carbon and silicon in the matrix are shown in Fig. 6a and 6b. These small grains could not be isotopically distinguished from one another considering the error bar; thus the isotopic values given in Table 3 stand for average values of the IOM in the image. Within error, the $^{15}N/^{14}N$ and $^{13}C/^{12}C$ ratios are the same for the IOM in all the different matrix areas.

On the NanoSIMS images of four large graphitized inclusions (with diameters ≥ 10 μm) in metallic nodules, no clear N or C isotopic hotspots were observed. Their mean isotopic compositions are given in Table 3 and the NanoSIMS images of the $^{15}N/^{14}N$ ratio and of the silicon distribution in one of the four graphitized inclusion are shown in Fig. 6c and 6d. This inclusion is also shown on the SEM picture Fig. 5c. Once again, the $^{15}N/^{14}N$ ratio of the inclusions is the same for the 4 inclusions.

The mean value of the $^{13}C/^{12}C$ ratio that we found for IOM in the matrix of SAH97096 (Table 3) is, within errors, in the range of values found by Deines and Wickman (1985) for EH4 to 6 and for IOM of EH3 to 4 in Alexander et al. (2007). Grady et al. (1986) analyzed EH4, EH5 and EL6 chondrites using stepped pyrolysis. They found a large range of $^{13}C/^{12}C$ ratio including the value we found for SAH97096 IOM and that disordered IOM (released at low temperature < 600°C) possesses a lower $^{13}C/^{12}C$ ratio than graphite (released at higher temperatures) –typical variation for $\delta^{13}C$ from -25 ‰ to 0 ‰.

The elemental H, C and N distributions were studied using the protocol described in §2.1.3. Fig. 7 contains the frequency histogram of distribution of the N/C ratio in the graphitized inclusions. The distribution of graphitized inclusions in metal shows an over abundance of N (right foot of the peak in Fig. 7), that demonstrates the presence of a N-rich component localized in metal. Moreover on the NanoSIMS images, we observed such N/C-rich zones as isolated inclusions in the metal (Fig. 6c and 6d). We noted that these zones are also Si-rich and do not contain C and N isotopic anomalies. They probably correspond to the silicon nitrides observed as inclusions in the metallic nodules of the EH3 Qingzhen (Alexander et al. 1994; Russell et al. 1995) and considered as precipitates exsolved during the metamorphism from the Si-bearing phases: kamacite, schreibersite or perryite.

### 3.3. Bulk and molecular structure of the IOM

The proportion of soluble and insoluble organic matter along with the inorganic content (in wt. %) and the relative H, C, N and F atomic fractions in the acid insoluble residues isolated from the EH3 SAH97096 are reported in Table 4. The soluble organic matter (SOM) constitutes less than 2% of the total organic matter (IOM and SOM) for both part of SAH97096. Therefore we will consider hereafter that the organic matter found *in situ* is IOM.

In addition to the standard HF/HCl treatment, the acid residue of SAH97096 isolated from the altered part, IOMa, has undergone an additional acid treatment ($HBO_3/HCl$). In this paragraph, we will detail the composition of this residue before and after the boron treatment. The inorganic matter constitutes around 53 and 49 wt.% of the acid residues, respectively, in the IOMa and IOMp and is mostly composed of indigenous oxides and sulfides known to survive the acid treatment together with fluorides formed during the acid treatment. Before the boron treatment, IOMa had a lower C content than IOMp (12% instead of 27%) and a much higher F content (up to 26% instead of 3.5%). After the boron treatment, the composition of IOMa reveals the efficiency of the $HBO_3/HCl$ treatment in reducing the fluoride content (F decreases from 26 wt.% to 2 wt.%) and increasing the proportion of carbonaceous matter in the residue (C increases from 12 to 37 wt.% C). The carbon concentration and the elemental H-C-N contents of the IOMp and IOMa (after $HBO_3$ treatment) are consistent with the values found in the literature for other enstatite chondrites (Grady et al. 1986; Alexander et al. 2007).

With Curie point pyrolysis/GC-MS, only a few types of aromatic compound were detected in the SAH97096 IOM as shown in the Fig. 8. Identical compounds are observed in the pyrolysate of IOMa and IOMp. These compounds are mostly aromatic (toluene, ethenylbenzene, naphthalene, phenol, 4-(methylthio)phenol, 3-methyl- and benzo- thiophene) and some contain sulphur in their aromatic cycles. The possibility that carboxylic compounds may be terrestrial contaminants cannot be ruled out even though they were found in both preparations.

EPR was performed on the IOM was performed to search for organic free radical or diradical species as observed in the IOM of Orgueil and Murchison (Binet et al. 2002, 2004b). On the EPR spectra (EPR intensity vs. magnetic field) of the carbonaceous chondrite IOM, two signals are superimposed: a broad feature (several hundreds of Gauss wide) from residual magnetic minerals in the IOM, which corresponds to a ferromagnetic resonance signal and a sharp signal of the organic radicals (only a few Gauss wide). For the 1 mg of SAH97096 IOM presently analyzed, only a broad signal (500 G wide) of superparamagnetic mineral nanoparticles appears on the EPR spectra (see Fig. 9a). Fig. 9b is an expanded view of the EPR spectra corresponding to the field range where the signal of the radicals in the IOM should appear. Since no signal of the radicals is observed for SAH97096 IOM, their lowest spin concentration is lower than ~$10^{13}$ radicals per gram.

Raman microspectroscopy was used to investigate the structural order degree of the organic matter by comparing relative intensity and width of the two main bands called G for "graphite band" and D for "defect band". The G region around 1600 cm$^{-1}$ includes the G band *sensu stricto* due to the stretching vibration of C atoms in aromatic planes and, if the carbon is not a perfect graphite, a defect band $D_2$ at ~1620 cm$^{-1}$. This $D_2$ band appears as a shoulder on the G band which might be as intense as the G band in very disordered CM chondrite IOM which results in a broad G massif centered at about 1610 cm$^{-1}$. The D band at ~1355 cm$^{-1}$ is the major band corresponding to structural defects within the graphitic network (*e.g.*, Tuinstra and Koenig 1970, Bény and Rouzaud 1985, Bernard et al. 2010). Its surface increases with the degree of disorder. The Raman spectra obtained in different locations in the IOM pressed in gold foils are composed of the two main massifs –bands G and D– having almost the same intensity (Fig. 10). Low intensity bands around 700 cm$^{-1}$ certainly correspond to oxides remaining in the IOM (Wang et al. 2004). The G band demonstrates for the aromatic nature of the IOM from SAH97096 inferred from aforementioned techniques, but the prominent defect bands reflect the highly disordered character of the aromatic structure. In SAH97096 IOM, the small variations in the relative intensities and shapes of the two bands could be due to the presence of a major component that dominates the Raman signal and mixed with particles having various structural forms.

HRTEM shows a wide structural heterogeneity at the nanometer scale. Disordered carbon – stacked polyaromatic units as observed in carbonaceous chondrites – is the main phase with variable fringe length from about 1 nm up to 10 nm (Fig. 11a; Le Guillou et al. 2011b). This observation is coherent with the averaged structure of the IOM recorded by Raman spectroscopy in the matrix. Moreover, particles showing various structural forms are distinguishable: graphite or highly graphitized lamellae of a few hundreds of nanometers observed along the 002 axis (Fig. 11b), nanoparticles of diamonds or sulfides coated by 1 to 3 layers of carbons (Le Guillou et al. 2011b) and 10-20 nm sized onion-like carbon. Moreover, one aggregate of amorphous-like carbonaceous spheres with diameters of a few hundreds of nanometers (Fig. 11c) was also observed. These latter look quite similar to the so-called nanoglobules observed by Garvie and Buseck (2004) and Nakamura-Messenger et al. (2006), but do not exhibit a hollow core. Graphite or highly graphitized lamellae probably correspond to the highly graphitized inclusions observed by Raman microspectroscopy in the metallic nodules.

Such a heterogeneity causes the semi-quantitative structural data of this IOM – that is the layer length, the interlayer spacing and the number of stacked layers – difficult to assess (in contrast see for example, Derenne et al. 2005 for Orgueil and Murchison or Remusat et al. 2008 for Kainsaz where more quantitative average structural information were obtained).

**3.4. Isotopic and elemental distributions of the IOM imaged with the NanoSIMS**

Fig. 6e, 6f, 6g and 6h shows four images for one NanoSIMS analysis of IOMa obtained during the first session of analyses. These images show: (1) the relatively homogeneity of the C/H ratio except in some micron-sized areas having higher C/H and C/N ratio that most probably correspond to areas with highly graphitized particles seen with HRTEM (Fig.6e), (2) the heterogeneity of the C/N ratio with small areas (1-2 μm) exhibiting very low ratios (< 5% of the mean value) (Fig. 6g), (3) the homogeneity of the D/H ratio (within ± 20 %)(Fig. 6f), (4) the heterogeneity of the $^{15}N/^{14}N$ ratio that exhibit "anomalies" *i.e.* areas as small as ~ 1 μm enriched or depleted in $^{15}N$ relative to the average value (Fig. 6h). The most nitrogen-rich

areas do not correspond to the $^{15}$N-anomalies. The non-altered sample (IOMp) does not show any significant differences with the aforementioned features.

Variations (standard deviation over the mean value of the image) of the H/C and N/C ratios within an image are reported in Table 5 for SAH97096 IOM, the type III Kerogen, Murchison and Orgueil IOM. SAH97096 IOM shows the highest variations for the H/C and N/C ratios. This high heterogeneity could be partly due to the presence of graphitized organic matter within the disordered IOM. Frequency histograms of the N/C distributions for SAH97096 IOM and for the IOM inclusions in metal are reported in Fig. 7. The data processing employed to quantify the H-, C- and N-variations within a NanoSIMS image is described in §2.1.3.

As for the graphitized inclusions in metal, the isolated IOM histogram has a tail toward the high N/C values. The tail accounts for the presence of a N-rich component mixed with the IOM: it is certainly the silicon nitride observed *in situ* that remains in the residue after acid treatment.

The mean D/H ratio for the pristine part, IOMp, [183 ± 10] x10$^{-6}$ ($\delta$D = + 179 ± 24 ‰) obtained over 5 images (~ 2000 µm$^2$) is significantly lower than for the altered part, IOMa, [206.30 ± 4.58] x10$^{-6}$ ($\delta$D = + 324 ± 56 ‰) obtained over 3 images (~ 1200 µm$^2$). On the other hand, the $^{15}$N/$^{14}$N ratios are similar in IOMp and IOMa when we consider 2$\sigma$ errors. The mean values for the fresh part [3.65 ± 0.04] x10$^{-3}$ ($\delta^{15}$N = -6 ± 11 ‰) and for the altered part is [3.68 ± 0.03] x10$^{-3}$ ($\delta^{15}$N = -0.2 ± 7.0 ‰) are identical to the value found *in situ* (Table 3). These isotopic ratios are in agreement with the data previously reported for enstatite chondrites (Grady et al. 1986 using stepped pyrolysis and combustion; Alexander et al. 2007 using elemental analysis).

We defined the isotopic hotspots or coldspots according to the definition given by Busemann et al. (2006): their sizes are of the same order as the spatial resolution and they have an isotopic ratio $R_{spot}$ such as: $|R_{spot} - R_{average}| > 3 \times \sigma_{spot}$. Following this definition, no deuterium hotspots were detectable, which is consistent with Fig. 6f. However, some $^{15}$N/$^{14}$N images reveal an occurrence of several coldspots and hotspots (Figs. 6d and 6e). The coldspots are rare: only 3 were detected over 12 NanoSIMS images of the IOM. Only one of the three was observed when Si was analysed (session #2). Its $^{15}$N/$^{14}$N ratio reaches [2.78 ± 0.115] x10$^{-3}$ ($\delta^{15}$N = -245 ± 32 ‰) and it is spatially correlated with the silicon signal (registered as $^{28}$Si$^-$). Hotspots are more frequent: 13 observed over the 12 images. Based on our analytical conditions (see §2.1.3.), the volume sputtered during the analysis is larger than the one in Busemann et al. (2006); nevertheless, the relative fractions of hotspots in Busemann et al. (2006) and in the present study being similar (*i.e.* ~ 0.1 vol.%), the hotspots concentration in SAH97096 can be estimated to be ≤ 0.1 vol.%.

Two groups could be differentiated among these hotspots as shown for example in the NanoSIMS image Fig. 6i: a moderately $^{15}$N-rich type ($\delta^{15}$N = 223 ± 27 ‰) spatially correlated with a silicon-rich zone (one occurrence is shown as an example with a black arrow in Fig. 6i and 6j) and visible in the image during 80 cycles and a highly $^{15}$N-rich type ($\delta^{15}$N = +829 ± 42 ‰ and $\delta^{15}$N = +1658 ± 82 ‰) not spatially correlated with the silicon-rich zone (cf. two occurrences shown with white arrows in Fig. 6i and 6j) and visible during the restricted number of 15 cycles. Note that the $\delta^{15}$N values given for the latter two hotspots are underestimated because their sizes are smaller than the size of the primary beam (around 300 nm). As a comparison, the $^{15}$N-hotspots found by Thomen et al. (2009) with similar NanoSIMS settings (images of 20 x 20 µm$^2$, 256 x 256 pixels, primary beam of 19 pA and ~ 1 min per cycle) in the IOM isolated from the carbonaceous chondrites Orgueil and Murchison exhibit lower $\delta^{15}$N values (maximum +480‰) with a minimal thickness corresponding to 60 cycles. Moreover, we observed a systematic N enrichment relative to C and H during the cycles in which the $^{15}$N enrichment is visible: the measured N/C ratio increases up to 50 % mostly due to the N-enhancement.

## 4. DISCUSSION

Results on the organic matter found in the enstatite chondrite SAH97096 are summarized with the aim of performing two comparisons: between the IOM from the altered and the pristine parts of SAH97096 to decipher the effects of terrestrial weathering and between the IOM from enstatite chondrites and from carbonaceous chondrites to infer a possible distribution of the IOM in the early solar system. Then, we will discuss on the possible carrier for the $^{15}$N-hotspots as well as the occurrences of carbonaceous inclusions in metal nodules.

### 4.1. Comparison between weathered and non-weathered part of SAH97096

It seems that terrestrial weathering did not transform the structure, the chemistry and the isotopic compositions of IOM in SAH97096. Indeed, comparing the IOM isolated from both parts of SAH97096 (IOMa to IOMp), we observed (1) similar nanometer scale structure, (2) similar H, C, N composition, (3) similar $^{15}$N/$^{14}$N bulk value, (4) a lack of deuterium to hydrogen ratio heterogeneity, (5) the occurrence of $^{15}$N-anomalies, (6) similar aromatic pyrolysates and (7) a lack of radical species. Nevertheless, two differences are notable. The first difference appears during IOM isolation. It was necessary to submit the weathered part to an additional acid treatment with boron acid to avoid the fluorides formed during the HF-HCl treatment. The presence of these fluorides is certainly linked to the mineralogy of the meteorite that changed due to the terrestrial oxidation and hydration of the meteorite. The altered minerals seem to react more easily with HF to form fluorides than minerals from the preserved fraction.

The other difference deals with the nature of the acid residue: the average D/H ratio for the weathered part is higher than for the preserved part. This difference cannot be explained by terrestrial contamination or isotopic exchange between terrestrial water and IOM as proposed by Ash and Pillinger (1995) for CR carbonaceous chondrites and ordinary chondrites (OC). If that were the case, the D/H ratio of the IOM from the altered part ($[206 \pm 5] \times 10^{-6}$) would have been closer to the terrestrial water D/H ratio ($156 \times 10^{-6}$) than to the one for the pristine part IOM ($[184 \pm 10] \times 10^{-6}$). By contrast, IOMp has higher F content and H/C ratio and lower C content than IOMa after boron treatment. This indicates that IOMp contains more inorganic residue than IOMa. These inorganic residues (containing fluorides) could contribute to the total budget of H in the IOMp, increasing the H/C ratio and decreasing the D/H ratio of IOMp.

The strong similarity between the altered and the non-altered part of SAH97096 contrasts with the observations of Ash and Pillinger (1995). These authors have observed that CR carbonaceous chondrites and ordinary chondrites found in the Sahara were depleted in light elements (C, N, H) by 21 to 45% relative to CR chondrites and ordinary chondrite falls or from Antarctic. They estimated up to 80 % of the organic material was lost due to the Saharan weathering. For SAH97096, the IOM loss between the non-altered and altered parts is 28 % (calculated from the wt. % IOM given in Table 4). This percentage is probably overestimated when taking into account the higher fluorine content in the non-altered part. The pronounced aromatic nature of the IOM in SAH97096 compared with the rather aliphatic-rich IOM in carbonaceous chondrites, is likely the cause for this selective preservation. According to Ash and Pillinger (1995), the loss of organic matter of the Saharan meteorites tends to decrease the bulk D/H ratio compared with other CR chondrites and ordinary chondrites. We observed the reverse tendency for SAH97096. Therefore the D/H ratio measured for the pristine part (IOMp) is probably an underestimation of the real D/H ratio of the IOM due to the contribution of terrestrial phases remaining in the acid residue.

**4.2. Comparison between SAH97096 and organic matter of carbonaceous chondrites**

Links between organic matter of enstatite and of carbonaceous chondrites, if any, could bring information on the distribution of organic grains in the early solar system. We now review differences and similarities between the IOM isolated from SAH97096 and the carbonaceous chondrites IOM.

In comparison with the IOM of the Orgueil and Murchison primitive carbonaceous chondrites, the IOM of SAH97096 exhibits differences in composition, in macromolecular structure, in degree of organization and in the isotopic anomalies.

Indeed, the pyrolysates of SAH97096 reveal the low aliphaticity of the macromolecule. They exhibit the limited number (~ 10) and nature of the pyrolysed organic molecules – mostly aromatic– and mark the difference with those of Orgueil and Murchison IOM, which show a diversity of aromatic and aliphatic products (Remusat et al. 2005). The H/C and N/C ratios of the SAH97096 IOM reach 0.310 and 0.005, respectively. In comparison with the IOM of the primitive carbonaceous chondrites Orgueil (H/C = 0.67 and N/C = 0.035) and Murchison (H/C = 0.59 and N/C = 0.033), SAH97096 IOM is clearly depleted in hydrogen and nitrogen relative to carbon. By contrast, the composition of SAH97096 IOM is in the range of values measured for CO and CV IOM (Alexander at al., 2007).

The SAH97096 IOM has a higher degree of organization than observed for primitive carbonaceous chondrites when comparing the G band and D band parameters on the Raman spectra (Quirico et al. 2011). Moreover, the fringe lengths (up to 10 nm) measured from the HRTEM pictures in the disordered material of SAH97096 IOM are on average longer than in Orgueil or Murchison IOM (Derenne et al., 2005).

Contrary to the IOM of primitive carbonaceous chondrites, we do not observe any deuterium hotspots. Furthermore no signal for radicals is observed in the EPR spectra for SAH97096 IOM. Radicals, if present in SAH97096 IOM, are almost five to six orders of magnitude below the radical concentrations in the IOM of carbonaceous chondrites.

The highest $^{15}$N-rich hotspots occurring in the SAH97096 IOM are distinct from those observed in carbonaceous chondrites. They are less abundant and less resistant under the primary ion beam of the NanoSIMS. They show high $^{15}$N-enrichment and higher N/C ratio than the mean IOM. A discussion on the possible carrier of these anomalies is given in a following section.

The major similarity between the IOM of both types of meteorites concerns its H, C, N global isotopic distribution. Indeed, the D/H and $^{15}$N/$^{14}$N ratios found for the isolated IOM and the $^{15}$N/$^{14}$N and $^{13}$C/$^{12}$C ratios found for IOM *in situ* for SAH97096 are in the range for carbonaceous chondrite IOM (*e.g.*, Geiss and Reeves 1981; Robert 2002; Alexander et al. 2007). In particular, comparison of the D/H and $^{15}$N/$^{14}$N ratios for SAH97096 and carbonaceous and ordinary chondrites (Fig. 12; data from Alexander et al. 2007) indicates that SAH97096 IOM is far from the domain defined by UOC and CR chondrites but lies in those of the CV-CO and CM-CI chondrites. More precisely, the D/H and $^{15}$N/$^{14}$N ratios of SAH97096 are similar to those defined by the thermally processed CO and CM chondrites. Moreover, the SAH97096 IOM is fairly similar to CO and CV chondrite IOM according to their D/H *vs*. H/C ratios and their $^{15}$N/$^{14}$N *vs*. C/N ratios (*cf.* Alexander et al. 2007). As SAH97096 is one of the least metamorphosed enstatite chondrites known, its D/H ratio is probably one of the least modified since accretion. Indeed, its H/C and D/H ratios have among the highest values found for enstatite chondrites (Alexander et al. 2007), which implies that the H/C and D/H lower ratios previously reported in the literature for IOM of enstatite chondrites (Alexander et al. 2007; Grady et al. 1986) should be regarded as resulting from secondary processes. A similar evolution is observed between primitive CM chondrites and heated CM chondrites (Alexander et al. 2007).

In addition to these isotopic features, both SAH97096 and a few carbonaceous chondrites exhibit disordered carbon located in the matrix that accounts for the major part of the IOM and highly graphitized organic matter present as inclusions within the metal nodules. *In situ* observations by Raman spectroscopy confirm the presence of IOM widespread within the matrix and between chondrules, chondrules fragments and opaque nodules (at the scale of the Raman mapping) as inferred from the single point Raman spectra of the matrix performed by Quirico et al. 2011. Homogeneous distribution of the IOM below the micron scale was observed in the matrix of CR, CM and CI chondrites by HRTEM and is likely a consequence of fluid circulation (Le Guillou et al. 2010, 2011a). In SAH97096, as in other enstatite chondrites, no evidence for liquid water circulation was observed so far. The homogeneous distribution of IOM observed by Raman spectroscopy in the whole rock SAH97096 suggests that such a distribution could predate the fluid circulation for carbonaceous and ordinary chondrites, but requires smaller scale investigation (as done for CR, CM and CI chondrite matrix).

If the isotopic and *in situ* location properties of the IOM in SAH97096 link the IOM of enstatite and carbonaceous chondrites, significant differences still exist. Are these differences inherited from different precursors for IOM in enstatite and carbonaceous chondrites or are they produced by secondary processes in the parent body? Based on different criteria (texture, enstatite-metal-CaS assemblage, composition of opaque assemblages etc.), Zhang et al. (1995) have proposed a thermal history for EH chondrites: (1) high temperature equilibration of chondrules, metal and sulfides (troilite – oldhamite) around 800-1000°C before accretion corresponding probably to the chondrule forming event, (2) thermal metamorphism on an onion like parent body (< 600 °C for the EH3), (3) a probable brecciation stage that induced a rapid cooling fixing the mineral composition and (4) possible supplementary superficial thermal event after formation and cooling of the parent body. The noticeable degree of organization of the SAH97096 IOM, together with the loss of heteroelements and the low aliphaticity suggest a thermal transformation of the IOM as already observed for the higher metamorphic grade CO 3.6 Kainsaz meteorite (C/H = 0.16) (Bonal et al. 2007; Remusat et al. 2008). As shown previously (Binet et al. 2002; Gourier et al. 2008), the deuterium hotspots in Orgueil are located in radical-rich regions. By NanoSIMS imaging, it was observed that the deuterium hotspots were absent in Kainsaz IOM (Remusat et al. 2008) and that the deuterium hotspots were thermally destroyed at 600 °C under vacuum in Murchison IOM (Remusat et al. 2009), while EPR showed that Kainzas IOM was strongly depleted in radicals, with concentrations being two orders of magnitude lower than in Orgueil and with no detectable diradicaloid. It is thus possible that, if any D/H ratio anomalies and radicals occurred in the IOM of SAH97096, they were destroyed by temperature. The structure of the IOM trapped in the matrix, observable with Raman spectroscopy and HRTEM, is irreversibly modified by thermal metamorphism (e.g., Deurbergue et al. 1987, Beyssac et al., 2002, 2003a; Bonal et al. 2006, 2007; Busemann et al. 2007; Quirico et al. 2011). In particular, Quirico et al. (2011) studied the structural order of a series of type 3 enstatite chondrites including SAH97096. In many respects, our study concurs with their observations, namely: the structural evolution of the enstatite chondrite IOM with metamorphism is comparable with similar observations in unequilibrated ordinary chondrites or in CO and CV carbonaceous chondrites; it probably results from the existence of a similar precursor for the IOM of these different meteorites, and the highly reduced environment does not influenced the IOM structural evolution.

To conclude, the SAH97096 IOM is in many respects comparable to the IOM of moderately metamorphosed carbonaceous chondrites as already suggested in preliminary studies (Piani et al. 2009, 2010). This observation is in agreement with the idea of a common precursor suggested by Alexander et al (1998) based on the abundance of presolar nanodiamonds in IOM. The following section addresses the specific question of the rare and small $^{15}$N-hotspots (sizes less than to 500 nm) with high N/C, apparently different from those in carbonaceous chondrites. These $^{15}$N-hotspots have survived the various thermal events linked to the history of the enstatite chondrites (Zhang et al. 1995).

## 4.3. What could be the carrier of the $^{15}$N-hotspots?

The organic matter found in meteorites as well as water and organic matter on Earth are systematically enriched in D and $^{15}$N relative to the solar values *i.e.* D/H = [20.0 ± 3.5] x10$^{-6}$ (δD = -872 ± 23 ‰) (Geiss and Gloeckler 2003) and $^{15}$N/$^{14}$N = [2.26 ± 0.67] x10$^{-3}$ (δ$^{15}$N = -375 ± 185 ‰) (Marty et al. 2010). Moreover, carbonaceous chondrite IOM contains D and/or $^{15}$N-rich anomalies (the so-called hotspots) at the micrometer scale. The $^{15}$N-enrichments of these hotspots range from δ$^{15}$N = 135 ‰ to 480 ‰ in Orgueil and Murchison (Thomen et al. 2009) and up to δ$^{15}$N ~2000 ‰ in CR chondrites (Floss and Stadermann 2009 and Busemann et al. 2006). They exceptionally reach δ$^{15}$N up to 3200 ‰ in the CM2 Bells (Busemann et al. 2006) and up to 4000 ‰ within clasts in the CB/CR Isheyevo (Briani et al. 2009; Bonal et al. 2010).

The carrier of the $^{15}$N-hotspots is rarely identified. Some of these $^{15}$N-hotspots could be related to organic nanoglobules observed in the IOM for the carbonaceous chondrite Tagish-Lake (Nakamura-Messenger et al. 2006). However, thanks to an HRTEM-NanoSIMS coordinated study by Abreu and Brearley (2010) of the matrix of the CR3.0 MET 00426, Le Guillou et al. (2010) and Floss et al. (2011) have observed that all $^{15}$N-hospots do not correspond to nanoglobules. Indeed, some hotspots are made of pseudo-amorphous homogeneous organic matter particles with vein shapes located between the silicate groundmass.

As in carbonaceous chondrites, the SAH97096 IOM is enriched in D and $^{15}$N relative to the solar values. Hotspots in SAH97096 IOM are either Si-rich with low δ$^{15}$N (less than 300 ‰) or Si-poor with high δ$^{15}$N (up to 1600 ‰). $^{15}$N-coldspots are rare and probably Si-rich (only one was observed in session #2). Neither hotpots nor coldspots have any deuterium enrichment compared to the mean value of the IOM (δD around 180 ‰) unlike the nanoglobules observed by Nakamura-Messenger et al. (2006). Recently, De Gregorio et al. (2010) reported nitrogen isotopic measurements for nanoglobules from Murchison and from particles of the Comet 81/Wild 2 of the Stardust collection having +289‰< δ$^{15}$N< +470 ‰ and up to +1120 ‰, respectively. The nanoglobule of Comet81/Wild 2 richest in $^{15}$N is the most aromatic and has N/C ratio from 0.1 to 0.15 significantly higher than the bulk IOM in comet particles and the IOM of primitive CM, CR and CI chondrites (Alexander et al. 2007). Hydrogen isotopic compositions for these nanoglobules appear to be uncorrelated with the $^{15}$N-enrichment. Nanoglobules from the comet Wild 2 and Si-poor $^{15}$N-hotspots in SAH97096 show similar properties: high $^{15}$N content, higher N/C ratio than the bulk IOM and no D-enrichment. In the course of this study, one aggregate of hundreds-of-nanometer-sized globules was observed on the HRTEM pictures in the SAH97096 IOM. It is therefore possible that similar nanoglobules exist in SAH97096 and in Stardust (De Gregorio et al. 2010).

Other carriers are possible for the $^{15}$N hotspots. Nittler (2003) reported three types of presolar grains that could be enriched in $^{15}$N: (1) silicon-carbide (SiC-X or SiC Nova type),

(2) graphite and (3) silicon-nitride ($Si_3N_4$). In the enstatite chondrites, the presence of silicon-carbides was reported in Qingzhen and Indarch (Huss and Lewis, 1995). Type X, B and mainstream SiC were also observed in SAH97166, paired with SAH97096 (Mahras et al., 2004). Presolar silicon-carbides are probably the carriers of the Si-rich hotspots and coldspots observed in our study. Nevertheless, the $^{15}$N-richest hotspots in SAH97096 contain no silicon and are depleted in carbon relative to nitrogen; none of the known presolar grains appears to be a promising candidate.

To the best of our knowledge, the carrier of these hotspots might correspond either to a type of nanoglobules as found by De Gregorio et al. (2010) in the Stardust particles or to a phase that has not been identified so far as the carrier of $^{15}$N-anomaly in other chondrites.

We have not been able to locate the N isotopic anomalies in whole-rock sample. This lack of detection may be only due to the low concentration of hotspots in the IOM and the low content of consistent IOM patches in matrix areas. Because it was easier to image large area of organic matter included in metal nodules and because this organic matter was isotopically homogeneous, we could assume that these $^{15}$N-hotspots are probably located within the matrix.

### 4.4. Inclusions in the metallic nodules of SAH97096

Inclusions of silicon nitride associated with organic matter were observed in the metal and could explain the C/N heterogeneities measured in the isolated IOM. Based on petrologic and thermodynamic calculations, Alexander et al. (1994) proposed that these euhedral nitrides (0.3 to 2.5 µm in size) could be exsolutions formed during the metamorphism on the EH3 Qingzhen and EH4 Indarch parent body. In contrast to the small size of nitrides, organic inclusions in kamacite nodules are often spheroidal and up to 30 µm in diameter. These inclusions possess a highly graphitized structure observable by Raman spectroscopy. According to the precision of the NanoSIMS analysis, the nitrogen and carbon (only two inclusions were measured for the $^{13}$C/$^{12}$C ratio) isotopic compositions are the same for all the inclusions. They also show the same $^{15}$N/$^{14}$N ratio as the IOM in matrix and a slight enrichment in $^{13}$C relative to this matrix IOM.

Organic matter inclusions in metal were first observed a long time ago in highly reduced meteorites such as iron meteorites (El Goresy 1965) or enstatite chondrites (Keil 1968). At present, they are also known to occur in the metal of primitive chondrites (carbonaceous and ordinary chondrites) (Mostefaoui et al. 1998, 2000, 2005) and primitive achondrites (El Goresy et al. 1995; Charon et al. 2009, 2010). Different processes are inferred for the formation of graphite: (1) direct nucleation and crystallization from carbon dissolved in a metal-carbon melt, (2) preliminary crystallisation of carbide (cohenite) that decomposed into metal and graphite (Brett 1967), (3) dissolution of the pre-existing IOM in liquid metal followed by the crystallisation of the carbon as graphite (Oya and Otani 1979) and (4) solid state exsolution through carbon diffusion.

In the case of enstatite chondrites, El Goresy et al. (1988) observed two types of graphite associated with the metal: graphite spherules in the Ni-poor kamacite of the EH3 and graphite veins in the Ni-rich kamacite of EH4. They proposed that graphite spherules could have formed by decomposition of cohenite during cooling of the metal (decreasing the Ni content of the metal by adding Fe) while the graphite veins would be exsolution products formed during metamorphism. Based on the elemental and isotopic variations between the inclusions from the same metal nodule or from different metal nodules, Mostefaoui et al. (2000) argued that these graphitized organic grains would have a homogeneous isotopic composition if they were formed from carbon dissolved in metal. They proposed that they could have formed from pre-existing heterogeneous IOM particles that were involved in the silicate reduction processes to a different extent.

The morphology of the carbon inclusions in kamacite (Fig. 5) is comparable with that observed by El Goresy et al. (1988) in EH3 chondrites, *i.e.* highly-graphitized with a spheroidal shape. This morphology does not suggest an exsolution process (no euhedral shape, no carbon in vein or in grain boundary) but, by contrast, spheroids of graphite are commonly produced in foundry material from liquid-state metallic alloy or by carbide decomposition (*e.g.*, Tartera et al. 2009).

Therefore the graphitized carbon inclusions likely formed through catalytic graphitization by the molten metallic alloy (direct nucleation or dissolution-rejection) involving a temperature above 1000°C or through the decomposition of cohenite involving a temperature around 600°C (Brett 1967). Consequently, their formation seems to be linked to high temperature event (such as chondrule formation, Zhang et al. 1995) incompatible with temperatures reached by the SAH97096 during its metamorphism.

The precursor of the carbon inclusions in metal could be the disordered matter found in the matrix. This latter suggestion is compatible with the similar N-isotopic compositions measured in the IOM of the matrix and in the metal inclusions and by the slight $^{13}$C enhancement in the graphitized carbon relative to the more disordered carbon in enstatite chondrite (Grady et al., 1986) that would result from evaporation/sublimation.

### 4.5. Concluding remarks

The IOM in the enstatite chondrite SAH97096 was studied using a battery of observational and analytical techniques. This study yielded information on different aspects of the organic matter distribution and evolution. A comparison between the two parts of the meteorite – a rusty part altered by terrestrial weathering and a well preserved part – shows that thermal fluctuations in the Sahara as well as oxidation did not affect the IOM structure, the H, C, N-chemistry, or the H and N isotopic composition. Interestingly, this observation raises the possibility of identifying extraterrestrial organic matter in terrestrial sedimentary rocks. Since all biologically derived insoluble organic matter on Earth show D/H values lower than SMOW (e.g. Schimmelmann et al., 2006), higher D/H ratios would signal the presence of preserved extraterrestrial IOM in sedimentary rocks.

In many respects, the IOM of SAH97096 could be considered as IOM from a moderately metamorphosed carbonaceous chondrite. The bulk H, N and C isotopic compositions are in the range of values found for carbonaceous and ordinary chondrite IOM. Several features of the SAH97096 IOM could be obtained by thermal metamorphism of an organic precursor having the properties of the primitive carbonaceous chondrite IOM, namely: the IOM structure shows a significantly higher degree of organization than for primitive carbonaceous chondrites, the IOM possesses higher C/N and C/H ratios, radicals and deuterium heterogeneities are missing. Although no trace of water circulation was observed in any enstatite chondrite, IOM in SAH97096 seems to be widespread within the matrix and between chondrules, chondrule fragments and opaque nodules, as in carbonaceous chondrites. Highly graphitized organic matter with spheroidal shape was found as inclusions in metal nodules. The nitrogen (and carbon) isotopic composition of these graphitized inclusions is the same as that of the matrix IOM indicating that it could be the result of a catalytic graphitization of the matrix IOM by the metal.

Occurrences of $^{15}$N-hotspots (rarer, smaller than observed in carbonaceous chondrites IOM but with high $^{15}$N-enrichments) could be inherited from a pre-accretion mixing between the chondritic organic precursor and a nitrogen rich source enriched in $^{15}$N. The carrier of these $^{15}$N-anomalies – potentially nanoglobules – has not been definitely identified.

**ACKNOWLEGMENT**

We gratefully thank the Muséum National d'Histoire Naturelle (Paris, France) and in particular B. Zanda for providing us samples of the SAH97096 for IOM isolation and resin-free section preparation. Christine Fiéni, Christelle Anquetil, Laurent Rémusat, Roger Hewins and Anders Meibom are thanked for their help in sample preparation, manuscript corrections and discussions. We are really grateful to A. Rubin and B. de Gregorio for their comments and reviews that helped us to clarify and improve the manuscript, and to associate editor Scott Sandford for careful editing. The Raman spectrometer at IMPMC was funded by an ANR JCJC Grant (project GeoCARBONS, PI O. Beyssac). This work has been supported by the Programme national de Planétologie (PNP INSU) and the ANR T-Tauri Chem.

**REFERENCES**

Abreu N.M. and Brearley A.J. 2010. Early solar system processes recorded in the matrices of two highly pristine CR3 carbonaceous chondrites, MET 00426 and QUE 99177. *Geochimica et Cosmochimica Acta* 74:1146–1171.

Alexander C.M.O'D., Swan P., and Prombo C.A. 1994. Occurence and implications of silicon nitride in enstatite chondrites. *Meteoritics and Planetary Science* 29:79–84.

Alexander C.M.O'D., Russell S.S., Arden J.W., Ash R.D., Grady M.M., and Pillinger C.T. 1998. The origin of chondritic macromolecular organic matter: a carbon and nitrogen isotope study. *Meteoritics and Planetary Science* 33:603–622.

Alexander C.M.O'D., Fogel M., Yabuta H., and Cody G. 2007. The origin and evolution of chondrites recorded in the elemental and isotopic compositions of their macromolecular organic matter. *Geochimica et Cosmochimica Acta* 71:4380–4403.

Alexander C.M.O'D., Newsome S.D., Fogel M.L., Nittler L.R., Busemann H., and Cody G.D. 2010. Deuterium enrichments in chondritic macromolecular materialimplications for the origin and evolution of organics, water and asteroids. *Geochimica et Cosmochimica Acta* 74:4417–4437.

Ammar, M. R. and Rouzaud, J.-N. 2011. How to obtain a reliable structural characterization of polished graphitized carbons by Raman microspectroscopy. J*ournal of Raman Spectroscopy*. doi: 10.1002/jrs.3014

Ash R. and Pillinger C. 1995. Carbon, nitrogen and hydrogen in Saharan chondrites: The importance of weathering. *Meteoritics and Planetary Science* 30:85–92.

Beny C., and Rouzaud, J. N.1985. Characterization of carbonaceous materials by correlated electron and optical microscopy and Raman microspectroscopy, in Scan. Electron. Micros., edited by S. E. M. Inc., v.12, pp. 119-132, AMF O'Hare, Chicago.

Bernard S., Beyssac O., and Benzerara K. 2008. Raman mapping using advanced line-scanning systems: Geological applications. *Applied Spectroscopy* 62:1180–1188.

Bernard S., Beyssac O., Benzerara K., Findling N., Tzvetkov G., & Brown Jr. G. 2010. XANES, Raman and XRD study of anthracene-based cokes and saccharose-based chars submitted to high-temperature pyrolysis. *Carbon* 48:2506–2516.

Beyssac O., Goffe B., Chopin, C. and Rouzaud J.N. 2002. Raman spectra of carbonaceous material from metasediments: a new geothermometer. *Journal of Metamorphic Geology*, 20, 859-871.

Beyssac O., Brunet F., Petitet J.P., Goffe B. and Rouzaud J.N. 2003a. Experimental study of the microtextural and structural transformations of carbonaceous materials under pressure and temperature. *European Journal of Mineralogy*, 15, 937-951.

Beyssac O., Goffé B., Petitet J., Froigneux E., Moreau M., and Rouzaud J. 2003b. On the characterization of disordered and heterogeneous carbonaceous materials using Raman spectroscopy. *Spectrochimica Acta, Part A* 59:2267–2276.

Binet L., Gourier D., Derenne S., and Robert F. 2002. Heterogeneous distribution of paramagnetic radicals in insoluble organic matter from the Orgueil and Murchison meteorites. *Geochimica et Cosmochimica Acta* 66:4177–4186.

Binet L., Gourier D., Derenne S., Pizzarello S., and Becke L. 2004a. Diradicaloids in the insoluble organic matter from the Tagish Lake meteorite: Comparison with the Orgueil and Murchison meteorites. *Meteoritics and Planetary Science* 39:1649–1654.

Binet L., Gourier D., Derenne S., Robert F., and Ciofini I. 2004b. Occurence of abundant diradicaloid moieties in the insoluble organic matter from the Orgueil and Murchison meteorites: a fingerprint of its extraterrestrial origin? *Geochimica et Cosmochimica Acta* 68:881–891.

Bonal L., Quirico E., Bourot-Denise M., and Montagnac G. 2006. Determination of the petrologic type of CV3 chondrites by Raman spectroscopy of included organic matter. *Geochimica et Cosmochimica Acta* 70:1849–1863.

Bonal L., Bourot-Denise M., Quirico E., Montagnac G., and Lewin E. 2007. Organic matter and metamorphic history of CO chondrites. *Geochimica et Cosmochimica Acta* 71:1605–1623.

Bonal L., Huss G.R., Krot A.N., and Nagashima K. 2010. Chondritic lithic clasts in the CB/CH-like meteorite Isheyevo: Fragments of previously unsampled parent bodies. *Geochimica et Cosmochimica Acta* 74:2500–2522.

Brett R. 1967. Cohenite: its occurrence and a proposed origin. *Geochimica et Cosmochimica Acta* 31:143–159.

Briani G., Gounelle M., Marrocchi Y., Mostefaoui S., Leroux H., Quirico E., and Meibom A. 2009. Pristine extraterrestrial material with unprecedented nitrogen isotopic variation. *Proceedings of the National Academy of Sciences* 106:10522–10527.

Busemann H., Young A.F., Alexander C.M.O'D., Hoppe P., Mukhopadhyay S., and Nittler L.R. 2006. Interstellar chemistry recorded in organic matter from primitive meteorites. *Science* 312:727–730.

Busemann H., Alexander C.M.O'D., and Nittler L.R. 2007. Characterization of insoluble organic matter in primitive meteorites by microraman spectroscopy. *Meteoritics and Planetary Science* 42:1387–1416.

Charon E., Aléon J., and Rouzaud J.N. 2009. Combined microraman and C- N- isotopic study of disordered carbons in Acapulcoites – Lodranites (abstract #5129). *Meteoritics and Planetary Science* 72:5159.

Charon E., Aléon J., and Rouzaud J.N. 2010. Determination of carbon origin in Acapulco and Lodran by HRTEM and C, N isotopes (abstract #5277). *Meteoritics and Planetary Science* 73:5277.

Cody G., Alexander C.M.O'D., Yabuta H., Kilcoyne A., Araki T., Ade H., Dera P., Fogel M., Militzer B., and Mysen B. 2008. Organic thermometry for chondritic parent bodies. *Earth and Planetary Science Letters* 272:446–455.

De Gregorio B.T., Stroud R.M., Nittler L.R., Alexander C.M.O'D., Kilcoyne A.L.D., and Zega T.J. 2010. Isotopic anomalies in organic nanoglobules from comet 81p/Wild 2: Comparison to Murchison nanoglobules and isotopic anomalies induced in terrestrial organics by electron irradiation. *Geochimica et Cosmochimica Acta* 74:4454–4470.

Deines P. and Wickman F.E. 1985. The stable carbon isotopes in enstatite chondrites and Cumberland falls. *Geochimica et Cosmochimica Acta* 49:89–95.

Derenne S., Rouzaud J.N., Clinard C., and Robert F. 2005. Size discontinuity between interstellar and chondritic aromatic structures: A high-resolution transmission electron microscopy study. *Geochimica et Cosmochimica Acta* 69:3911–3917.

Derenne S. and Robert F. 2010. Model of molecular structure of the insoluble organic matter isolated from Murchison meteorite. *Meteoritics and Planetary Science* 45:1461–1475.

Deurbergue A., Oberlin A., Oh J. and Rouzaud J.N. 1987. Graphitization of Korean anthracites as studied by transmission electron microscopy and X-ray diffraction. *International Journal of Coal Geology*, 8, 375-393.

Duprat J., Dobrica E., Engrand C., Aleon J., Marrocchi Y., Mostefaoui S., Meibom A., Leroux H., Rouzaud J.N., Gounelle M., and Robert F. 2010. Extreme deuterium excesses in ultracarbonaceous micrometeorites from central antarctic snow. *Science* 328:742–745.

El Goresy A. 1965. Mineralbestand und Strukturen der Graphit- und Sulfideinschlüsse in Eisenmeteoriten. *Geochimica et Cosmochimica Acta* 29:1131–1136, IN1–IN18, 1137–1151.

El Goresy A., Yabuki H., Ehlers K., Woolum D., and Pernick E. 1988. Qingzhen and Yamato-691: A tentative alphabet for the EH chondrites. *Antarctic Meteorite Research* 1:65–101.

El Goresy A., Zinner E., and Marti K. 1995. Survival of isotopically heterogeneous graphite in a differentiated meteorite. *Nature* 373:496–499.

Fletcher, I. R.; Kilburn, M. R. & Rasmussen, B. 2008. NanoSIMS µm-scale in situ measurement of 13C/12C in early Precambrian organic matter, with permil precision. *International Journal of Mass Spectrometry*, 278, 59-68.

Floss C. and Stadermann F.J. 2009. High abundances of circumstellar and interstellar C-anomalous phases in the primitive CR3 chondrites QUE 99177 and MET 00426. *The Astrophysical Journal* 697:1242–1255.

Floss C., Le Guillou C., Stadermann F.J., and Brearley A.J. 2011. Coordinated NanoSIMS and TEM analyses of C- and N-anomalous phases in the CR3 chondrites MET 00642 (abstract #1455). 42$^{nd}$ Lunar and Planetary Science Conference. CD-ROM.

Gardinier A., Derenne S., Robert F., Behar F., Largeau C., and Maquet J. 2000. Solid state CP/¨MAS $^{13}$C NMR of the insoluble organic matter of the Orgueil and Murchison meteorites: quantitative study. *Earth and Planetary Science Letters* 184:9–21.

Garvie L.A. and Buseck P.R. 2004. Nanosized carbon-rich grains in carbonaceous chondrite meteorites *Earth and Planetary Science Letters* 224:431 – 439.

Geiss J. and Gloeckler G. 2003. Isotopic composition of H, He and Ne in the protosolar cloud. *Space Science Reviews* 106:3–18.

Geiss J. and Reeves H. 1981. Deuterium in the solar system. *Astronomy and Astrophysics* 93:189–199.

Gourier D., Robert F., Delpoux O., Binet L., Vezin H., Moissette A., and Derenne S. 2008. Extreme deuterium enrichment of organic radicals in the Orgueil meteorite: Revisiting the interstellar interpretation? *Geochimica et Cosmochimica Acta* 72:1914–1923.

Grady M.M., Wright I.P., Carr L.P., and Pillinger C.T. 1986. Compositional differences in enstatite chondrites based on carbon and nitrogen stable isotope measurements. *Geochimica et Cosmochimica Acta* 50:2799–2813.

Hanon P., Robert F., and Chaussidon M. 1998. High carbon concentrations in meteoritic chondrules: A record of metal-silicate differentiation. *Geochimica et Cosmochimica Acta* 62:903–913.

Huss G. R. and Lewis R. S. 1995. Presolar diamond, SiC, and graphite in primitive chondrites: Abundances as a function of meteorite class and petrologic type. *Geochimica et Cosmochimica Acta*, 59:115–160.

Keil K. 1968. Mineralogical and chemical relationships among enstatite chondrites. *Journal of Geophysical Research* 73:6945–6976.

Kerridge J.F., Chang S., and Shipp R. 1987. Isotopic characterisation of kerogen-like material in the Murchison carbonaceous chondrite. *Geochimica et Cosmochimica Acta* 51:2527–2540.

Le Guillou C., Brearley A.J., Floss C., and Stadermann F. 2010. *In situ* observation of C and N anomalous organic grains in the matrix of MET 00426 (CR3.0)(abstract #5342). *Meteoritics and Planetary Science* 73:5342.

Le Guillou C., Remusat L., Bernard S., and Brearley A.J. 2011a. Redistribution and evolution of organics during aqueous alteration: Nanosims-SXTM-TEM analyses of FIB sections from Renazzo, Murchison and Orgueil (abstract #1996). 42$^{nd}$ Lunar and Planetary Science Conference. CD-ROM.

Le Guillou C., Rouzaud J.N., Bonal L., Quirico E., and Derenne S. 2011b. High resolution TEM of chondritic carbonaceous matter: metamorphic evolution and heterogeneity. *Meteoritics and Planetary Science* Submitted.

Marhas K. K., Hoppe P., Besmehn A. and E. Stansbery, S. M. 2004. A NanoSIMS Study of Iron-Isotopic Compositions in Presolar Silicon Carbide Grains (abstract #1834). 35$^{th}$ Lunar and Planetary Institute Science Conference Abstracts. CD-ROM.

Marty B., Zimmermann L., Burnard P.G., Wieler R., Heber V.S., Burnett D.L., Wiens R.C., and Bochsler P. 2010. Nitrogen isotopes in the recent solar wind from the analysis of genesis targets: Evidence for large scale isotope heterogeneity in the early solar system. *Geochimica et Cosmochimica Acta* 74:340–355.

Mostefaoui S., Hoppe P., and El Goresy A. 1998. *In situ* discovery of graphite with interstellar isotopic signatures in a chondrule-free clast in an l3 chondrite. *Science.* 280:1418–1420.

Mostefaoui S., Perron C., Zinner E., and Sagon G. 2000. Metal-associated carbon in primitive chondrites: Structure, isotopic composition, and origin. *Geochimica et Cosmochimica Acta* 64:1945–1964.

Mostefaoui S., Zinner E., Hoppe P., Stadermann F.J., and El Goresy A. 2005. *In situ* survey of graphite in unequilibrated chondrites: Morphologies, C, N, O, and H isotopic ratios. *Meteoritics and Planetary Science* 40:721–743.

Nakamura-Messenger K., Messenger S., Keller L.P., Clemett S.J., and Zolensky M.E. 2006. Organic Globules in the Tagish Lake Meteorite: Remnants of the Protosolar Disk. *Science* 314:1439–1442.

Nittler L.R. 2003. Presolar stardust in meteorites: recent advances and scientific frontiers. *Earth and Planetary Science Letters* 209:259–273.

Oya A. and Otani S. 1979. Catalytic graphitization of carbons by various metals. *Carbon* 17:131–137.

Piani L., Derenne S., Robert F., Thomen A., Mostefaoui S., Marrocchi Y. and Meibom A. 2009. Molecular and Isotopic Study of the Insoluble Organic Matter Isolated from a Primitive Enstatite Chondrite. (abstract #5134) *Meteoritics and Planetary Science*, 72.

Piani L., Robert F., Derenne S., Thomen A., Bourot-Denise M., Mostefaoui S., Marrocchi Y. and Meibom A. 2010. The Organic Matter in the Less Metamorphized Enstatite Chondrite Sahara 97096: Isotopic Composition and Spatial Distribution. (abstract #1736). 41st Lunar and Planetary Institute Science Conference. CD-ROM.

Quirico E., Montagnac G., Rouzaud J.N., Bonal L., Bourot-Denise M., Duber S., and Reynard B. 2009. Precursor and metamorphic condition effects on raman spectra of poorly ordered carbonaceous matter in chondrites and coals. *Earth and Planetary Science Letters,* 287, 185-193.

Quirico E., Bourot-Denise M., Robin C., Montagnac G., and Beck P. 2011. A reappraisal of the metamorphic history of EH3 and EL3 enstatite chondrites. *Geochimica et Cosmochimica Acta*, 75, 3088-3102.

Remusat L., Derenne S., Robert F., and Knicker H. 2005. New pyrolytic and spectroscopic data on Orgueil and Murchison insoluble organic matter: A different origin than soluble? *Geochimica et Cosmochimica Acta,* 69, 3919-3932.

Remusat L., Palhol F., Robert F., Derenne S., and France-Lanord C. 2006. Enrichment of deuterium in insoluble organic matter from primitive meteorites: A solar system origin? *Earth and Planetary Science Letters* 243:15 – 25.

Remusat L., Robert F., and Derenne S. 2007. The insoluble organic matter in carbonaceous chondrites: Chemical structure, isotopic composition and origin. *Comptes Rendus Geoscience* 339:895 – 906.

Remusat L., Le Guillou C., Rouzaud J.N., Binet L., Derenne S., and Robert F. 2008. Molecular study of insoluble organic matter in Kainsaz CO3 carbonaceous chondrite: Comparison with CI and CM IOM. *Meteoritics and Planetary Science* 43:1099–1111.

Remusat L., Robert F., Meibom A., Mostefaoui S., Delpoux O., Binet L., Gourier D., and Derenne S. 2009. Proto-planetary disk chemistry recorded by D-rich organic radicals in carbonaceous chondrites. *The Astrophysical Journal* 698:2087–2092.

Remusat L., Guan Y., Wang Y., and Eiler J.M. 2010. Accretion and preservation of D-rich organic particles in carbonaceous chondrites: Evidence for important transport in the early solar system nebula. *The Astrophysical Journal* 713:1048–1058.

Robert F. and Epstein S. 1982. The concentration and isotopic composition of hydrogen, carbon and nitrogen in carbonaceous meteorites. *Geochimica et Cosmochimica Acta* 46:81–95.

Robert F. 2002. Water and organic matter D/H ratios in the solar system: a record of an early irradiation of the nebula? *Planetary and Space Science* 50:1227 – 1234.

Rubin A.E. and Keil K. 1983. Mineralogy and petrology of the Abee enstatite chondrite breccia and its dark inclusions. *Earth and Planetary Science Letters* 62:118–131.

Rubin A.E. and Scott W.R. 1997. Abee and related EH chondrite impact-melt breccias. *Geochimica et Cosmochimica Acta* 61:425–435.

Rubin A.E., Griset C.D., Choi B.G., and Wasson J.T. 2009. Clastic matrix in EH3 chondrites. *Meteoritics and Planetary Science* 44:589–601.

Russell S.S., Lee M.R., Arden J.W., and Pillinger C.T. 1995. The isotopic composition and origins of silicon nitride from ordinary and enstatite chondrites. *Meteoritics and Planetary Science* 30:399-404.

Schimmelmann A., Sessions A. L., & Mastalerz M. 2006. Hydrogen Isotopic (D/H) Composition of Organic Matter During Diagenesis and Thermal Maturation. *Annual Review of Earth and Planetary Sciences*, 34, 501-533.

Tartera J., Marsal M., Varela-Castro G., and Ochoa de Zabalegui E. 2009. Looking at graphite spheroids. *International Journal of Metalcasting* 3:7–17.

Thomen A., Robert F., Mostefaoui S., Piani L., Marrocchi Y., and Meibom A. 2009. Spatial relations between D/H and N isotopic anomalies in orgueil and murchison insoluble organic matter: A nanosims study (abstract #5284). *Meteoritics and Planetary Science* 72:5284.

Thomen, A.; Remusat, L.; Robert, F.; Meibom, A. and Mostefaoui, S. 2010. Chemical and Nitrogen Isotopic Composition of the Hotspots in Orgueil Insoluble Organic Matter (abstract #2472). 41st Lunar and Planetary Institute Science Conference. CD-ROM.

Tuinstra, F. and Koenig, J. L. 1970. Raman Spectrum of Graphite. Journal of Chemical Physics 53, 1126-1130.

Wang A., Kuebler K.E., Jolliff B.L., and Haskin L.A. 2004. Mineralogy of a Martian meteorite as determined by Raman spectroscopy. *Journal of Raman spectroscopy,* 35:504-514.

Weisberg M.K. and Prinz M. 1998. Sahara 97096: A highly primitive EH3 chondrite with layered sulfide-metal-rich chondrules (abstract #1741). 29[th] Lunar and Planetary Science Conference. CD-ROM.

Yabuta H., Williams L.B., Cody G.D., Alexander C.M.O'D., and Pizzarello S. 2007. The insoluble carbonaceous material of CM chondrites: A possible source of discrete organic compounds under hydrothermal conditions. *Meteoritics and Planetary Science* 42:37–48.

Yang J. and Epstein S. 1983. Interstellar organic matter in meteorites. *Geochimica et Cosmochimica Acta* 47(12):2199–2216.

Zhang Y., Benoit P.H., and Sears D.W.G. 1995. The classification and complex thermal history of the enstatite chondrites. *Journal of Geophysical Research* 100:9417–9438.

# TABLES

Table 1. Carbon abundance measured by EPMA in three NIST reference materials and in 22 Fe-Ni metal grains in Sahara 97096.

| Sample | Wt.% C [a] | Wt.% C measured | Standard Deviation | N [b] |
|---|---|---|---|---|
| SRM 665 | 0.008 | 0.82 | ± 0.18 | 16 |
| SRM 661 | 0.39 | 1.43 | ± 0.27 | 4 |
| SRM 664 | 0.87 | 1.57 | ± 0.52 | 12 |
| Kamacite | | 0.76 | ± 0.09 | 22 |

(a) from NIST database
(b) N stands for the number of measurements

Table 2. Reproducibility (Repro) at 2σ (*i.e.* twice the ratio of the standard deviation to the mean of the N measured isotopic ratios) and Instrumental Mass Fractionation (IMF) for the type III kerogen used as a standard for the NanoSIMS imaging; IMF = $R_{Measured} / R_{known}$ (with R the isotopic ratio).

| Session | $D^-/H^-$ | | | $C^{15}N^-/C^{14}N^-$ | | | $^{13}C^-/^{12}C^-$ | | |
|---|---|---|---|---|---|---|---|---|---|
| | Repro | IMF | N | Repro | IMF | N | Repro | IMF | N |
| #1 | ± 4.1 % | 0.75 | 17 | ± 3.7 % | 1.02 | 17 | --- | --- | --- |
| #2 | --- | --- | --- | ± 1.6 % | 1.00 | 25 | ± 1.9 % | 1.03 | 8 |

Table 3. N- and C-isotopic ratios of the organic matter measured in the matrix and in the metal (the number of the data stands for the number of ion image). The error bars are given at 2σ.

| Organic matter | $^{15}N/^{14}N$ (x$10^{-3}$) | $^{13}C/^{12}C$ (x$10^{-3}$) |
|---|---|---|
| in matrix | | |
| 1- | 3.69 ± 0.15 | |
| 2- | 3.72 ± 0.09 | |
| 3- | 3.72 ± 0.11 | 11.13 ± 0.23 |
| 4- | 3.69 ± 0.11 | 10.94 ± 0.12 |
| **Average** | **3.70 ± 0.05** | **11.04 ± 0.13** |
| | ($\delta^{15}N$= 6 ± 11 ‰) | ($\delta^{13}C$= -17 ± 12 ‰) |
| in metal | | |
| 1- | 3.66 ± 0.08 | |
| 2- | 3.67 ± 0.07 | |
| 3- | 3.75 ± 0.07 | |
| 4- | 3.77 ± 0.09 | 11.18 ± 0.05 |
| 5- | 3.73 ± 0.10 | 11.24 ± 0.06 |
| **Average** | **3.72 ± 0.04** | **11.21 ± 0.04** |
| | ($\delta^{15}N$= 12 ± 11 ‰) | ($\delta^{13}C$= -2 ± 4 ‰) |

Table 4. Chemical analyses of SAH97096 for the bulk sample (columns 1-2) and for the HF-HCl residues (columns 3-7). In successive columns: (1) the concentration of the soluble organic matter (SOM); (2) the concentration of the insoluble organic matter (IOM; for the altered part the value was obtained before and after the $HBO_3$ treatment); (3-4-5) the carbon, fluorine and ash; (6-7) the H/C and N/C atomic ratios.

| Sahara 97096 | Whole Rock | | Acid residues | | | | |
|---|---|---|---|---|---|---|---|
| | SOM (wt.%) | IOM (wt.%) | C (wt.%) | F (wt.%) | Ash content (wt.%) | H/C at. | N/C at. |
| Pristine part | 0.011 | 0.603 | 27.2 | 3.5 | 49.0 | 0.311 | 0.005 |
| Altered part before $HBO_3$ | 0.006 | 1.223 | 11.8 | 26.5 | 53.3 | 1.723 | *N under d.l.* |
| Altered part after $HBO_3$ | - | 0.434 | 36.6 | 2.0 | n.d. | 0.180 | *N under d.l.* |

d.l. = detection limit, n.d.= not determined

Table 5. $H^-$, $N^-$ and $C^-$ relative variation within NanoSIMS images for SAH97096 IOM, the type III Kerogen, Orgueil and Murchison IOM : the reported values correspond to the standard deviation over the mean of the H/C and N/C ionic ratios over the ROI defined in the images (see §2.1.3. for details). They permit us to compare the heterogeneity of the H/C and N/C ionic ratios in various samples.

| | $H^-/{}^{12}C^-$ | ${}^{12}C^{14}N^-/{}^{12}C^-$ |
|---|---|---|
| SAH97096 IOM | 26.7 % | 60.0 % |
| Kerogen III | 12.4 % | 23.4 % |
| Orgueil IOM | 9.7 % | 17.6 % |
| Murchison IOM | 8.4 % | 21.3 % |

FIGURES

Fig. 1.

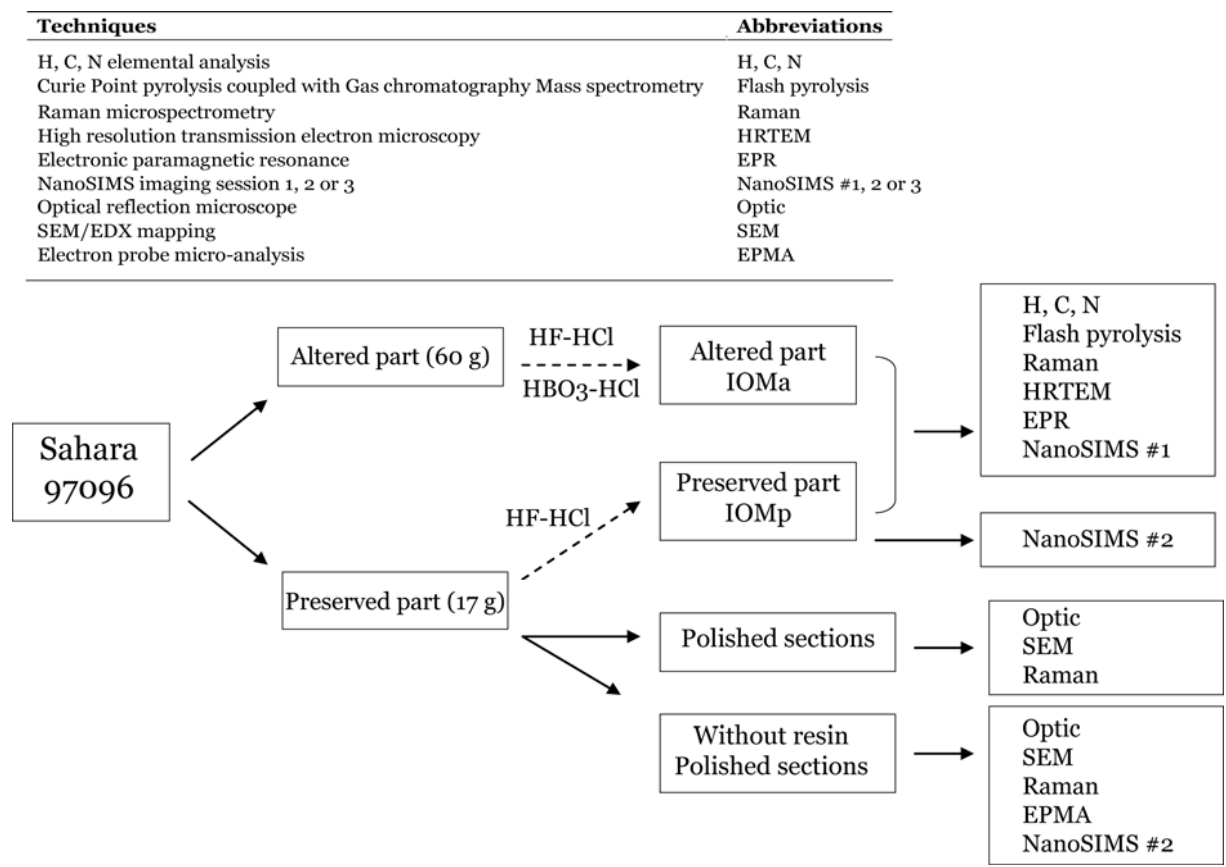

Fig. 2.

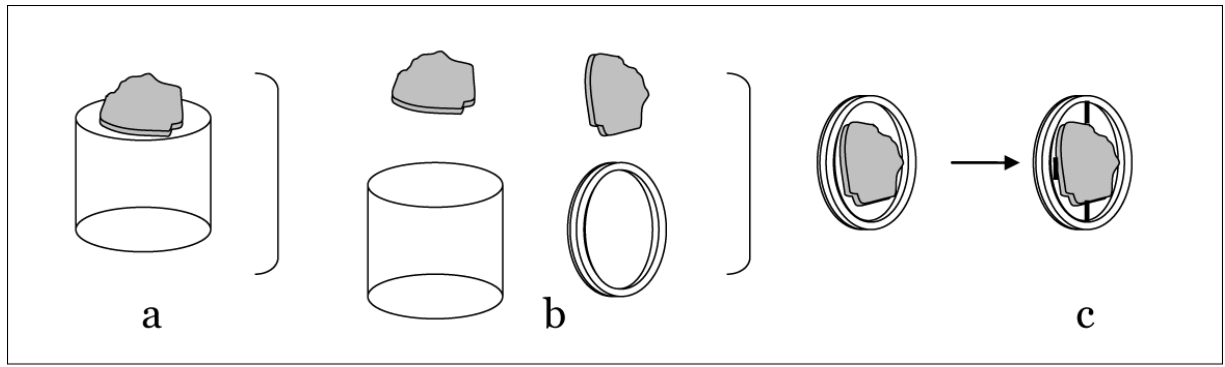

Fig. 3.

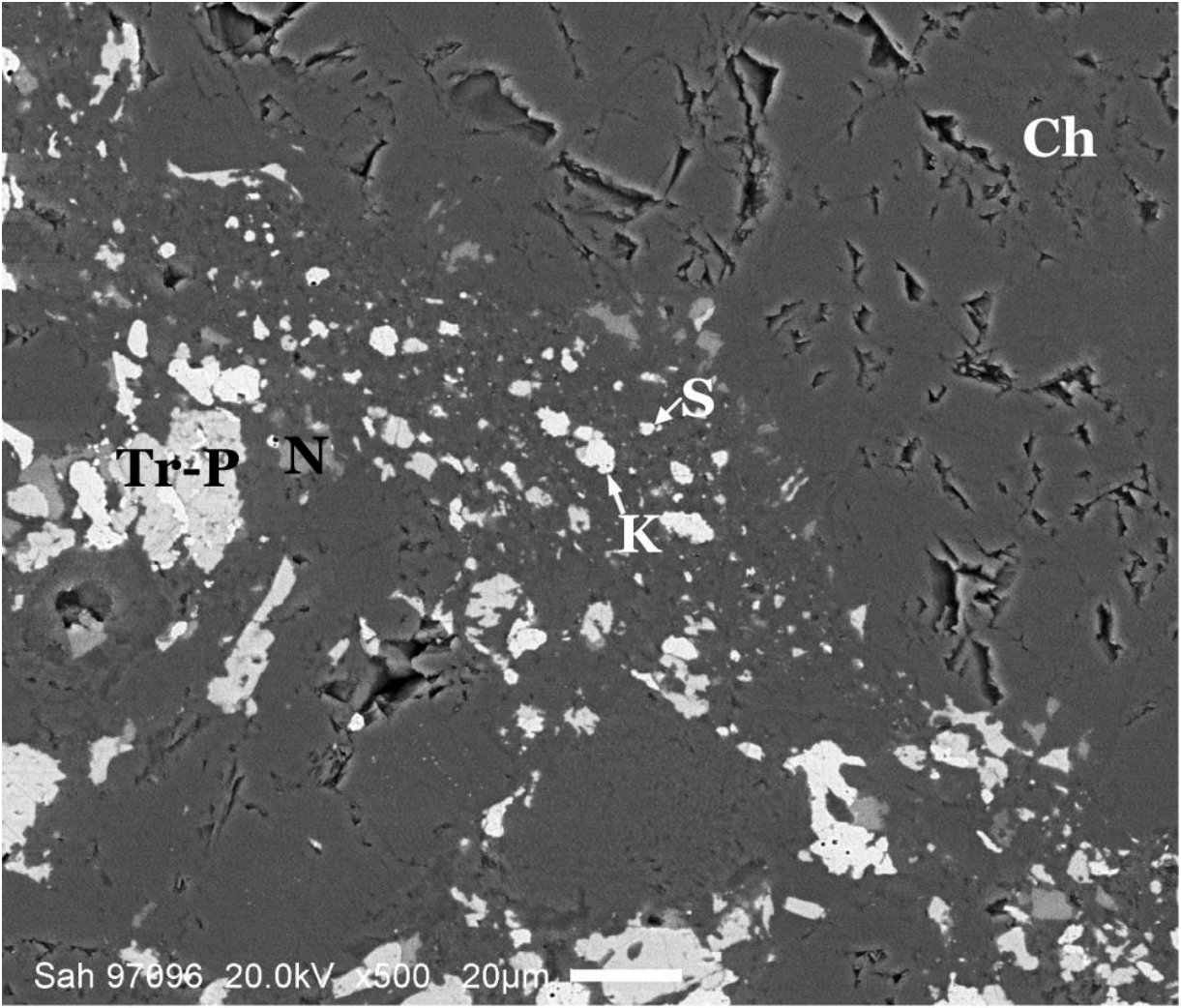

Fig. 4.

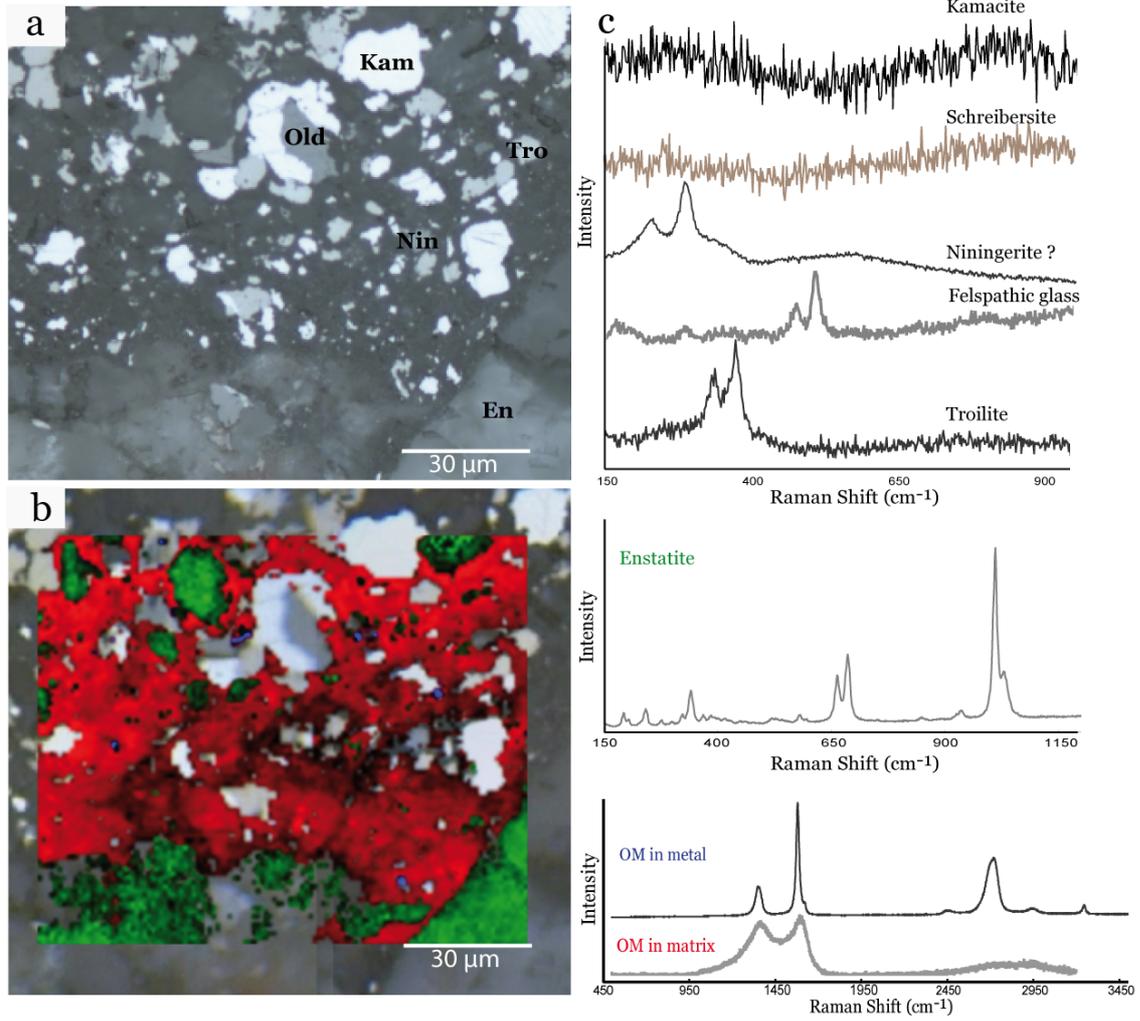

Fig. 5.

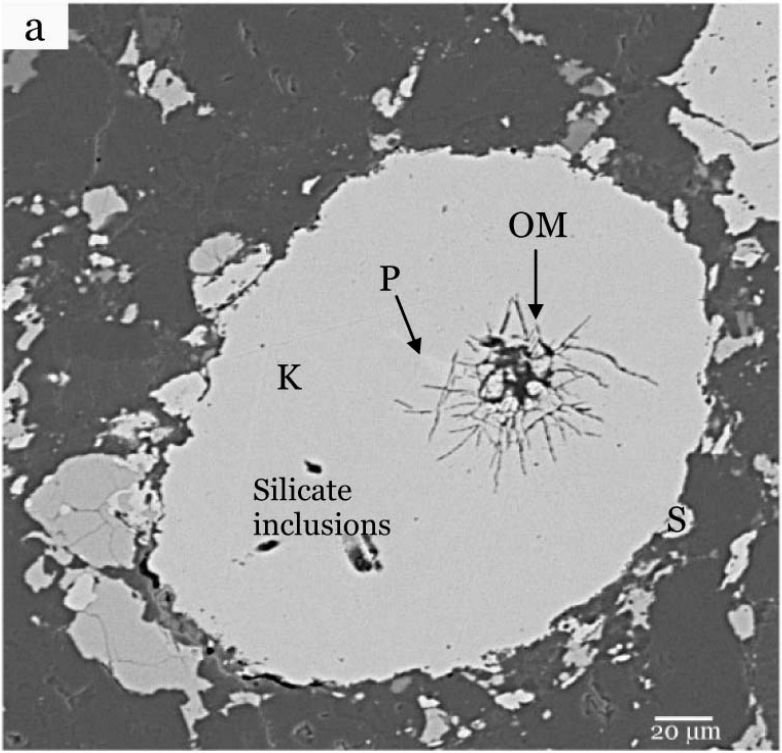 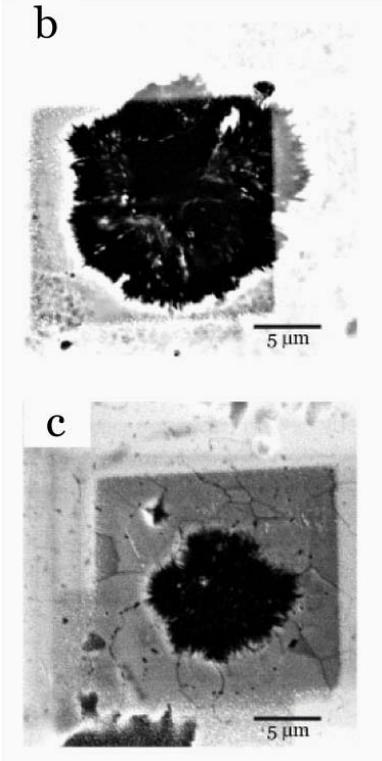

Fig. 6.

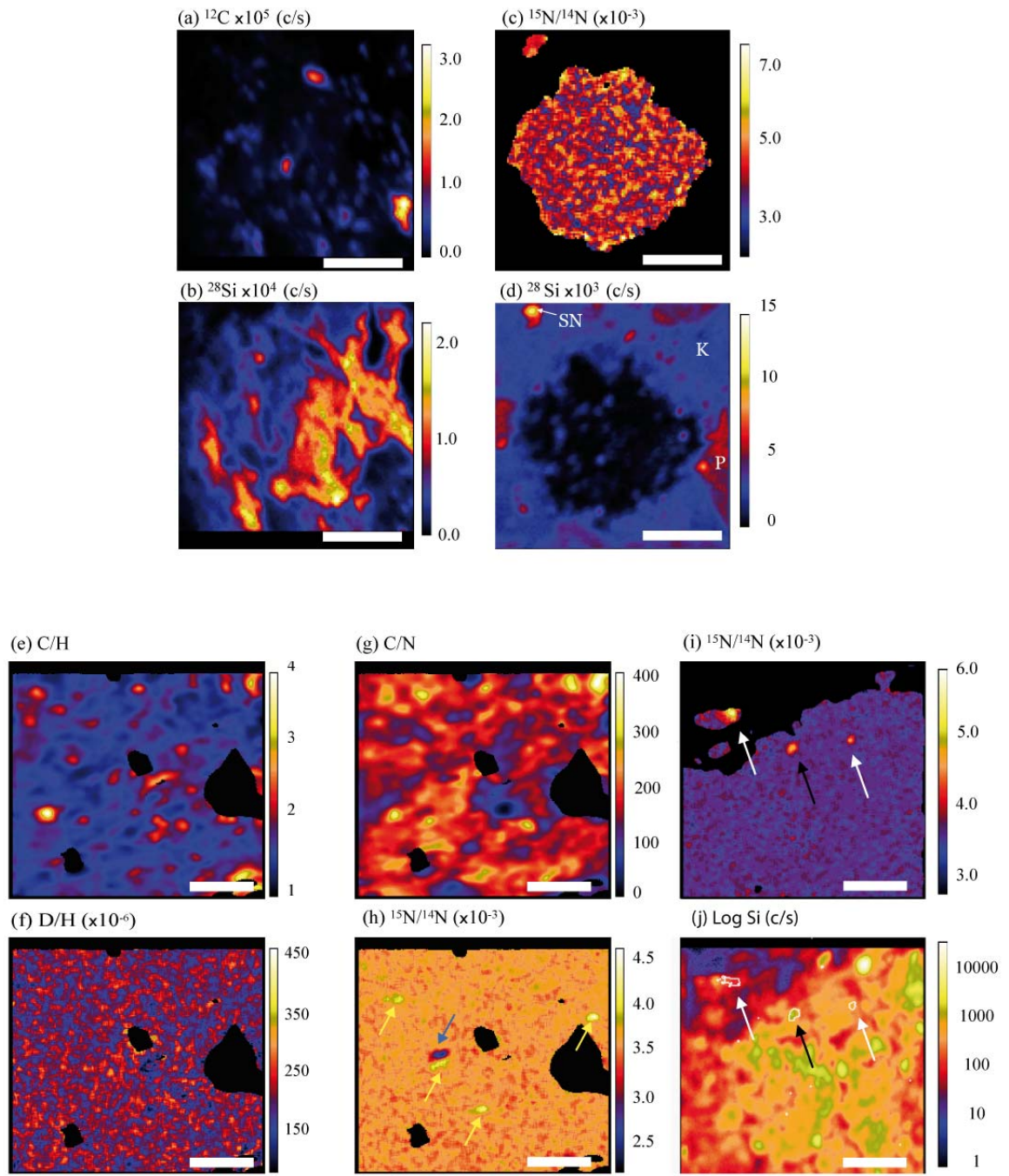

Fig. 7.

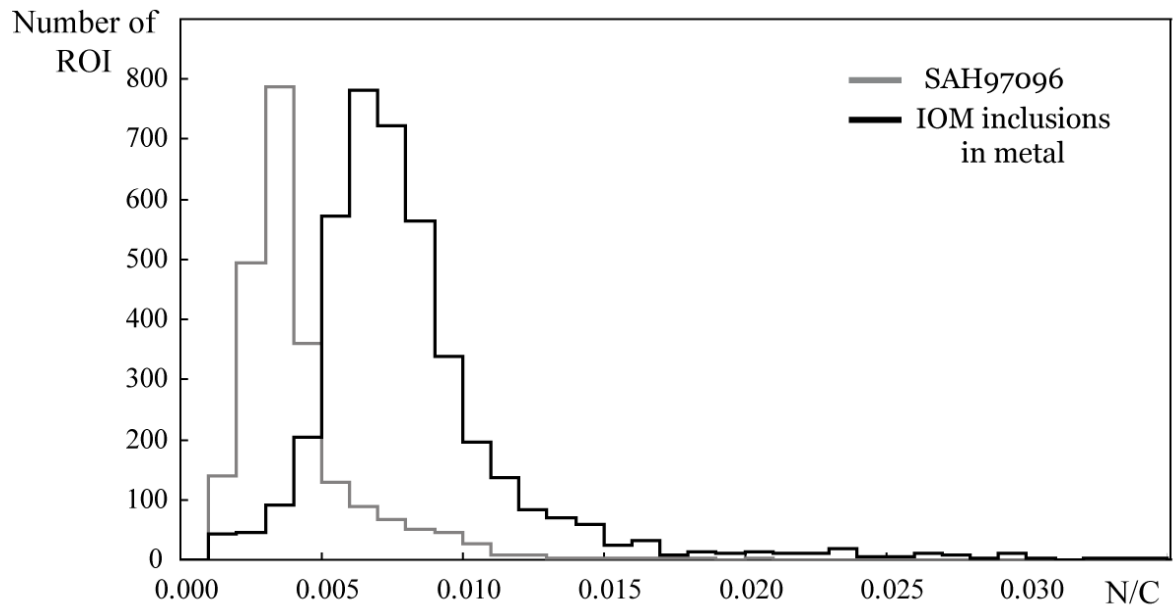

Fig. 8.

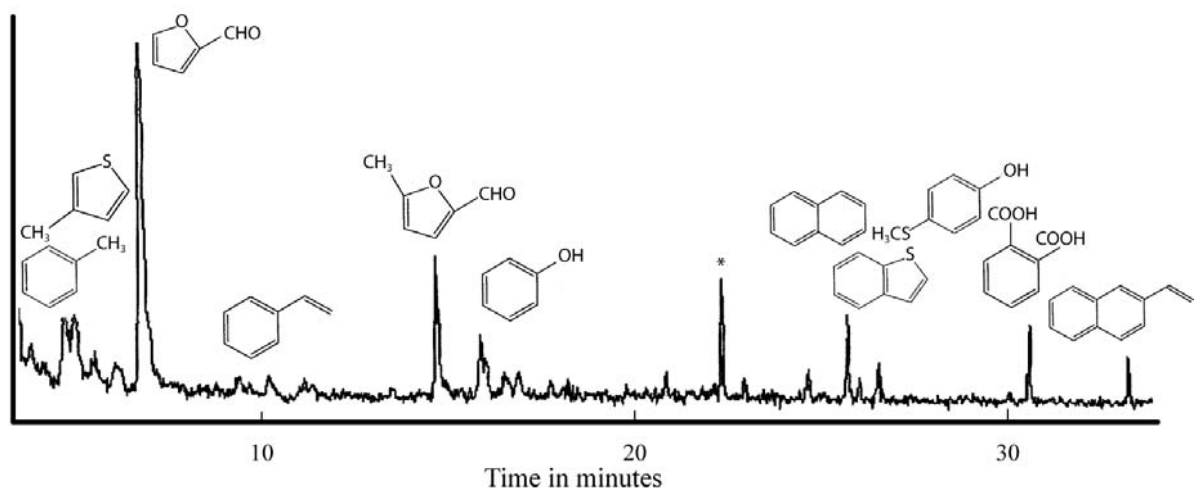

Fig. 9.

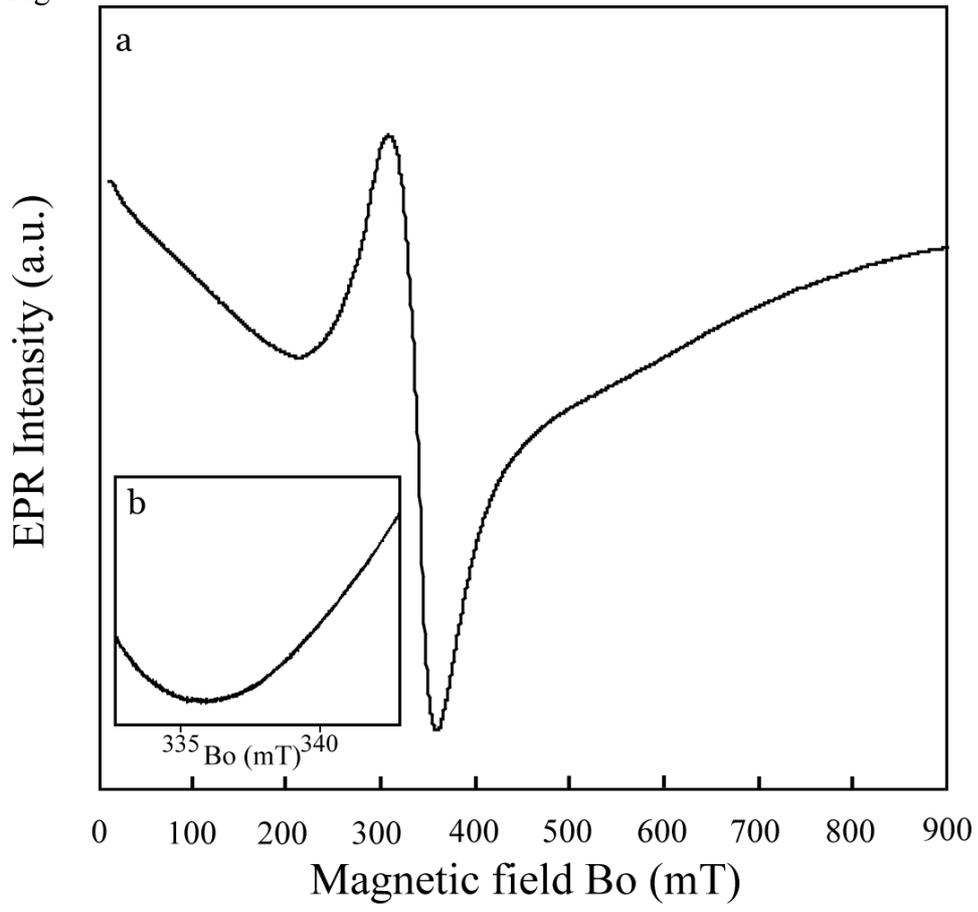

Fig. 10.

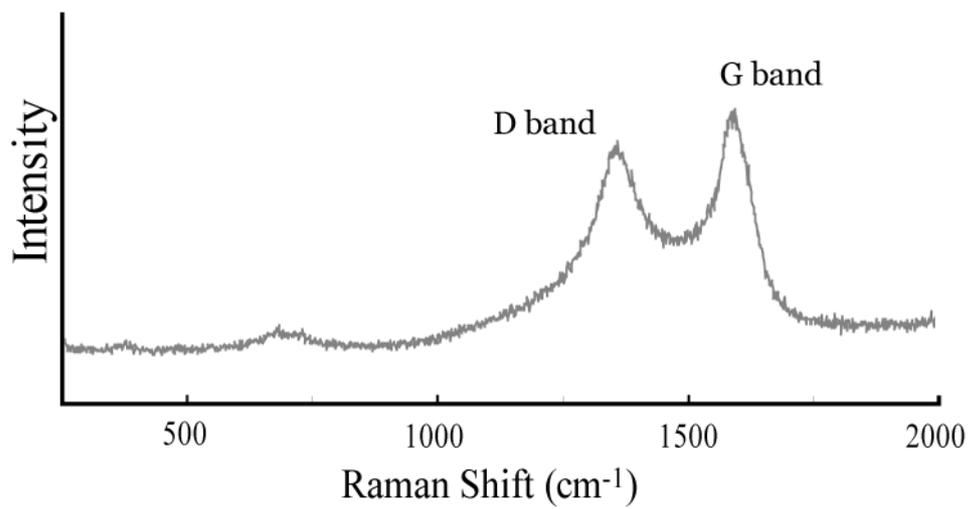

Fig. 11.

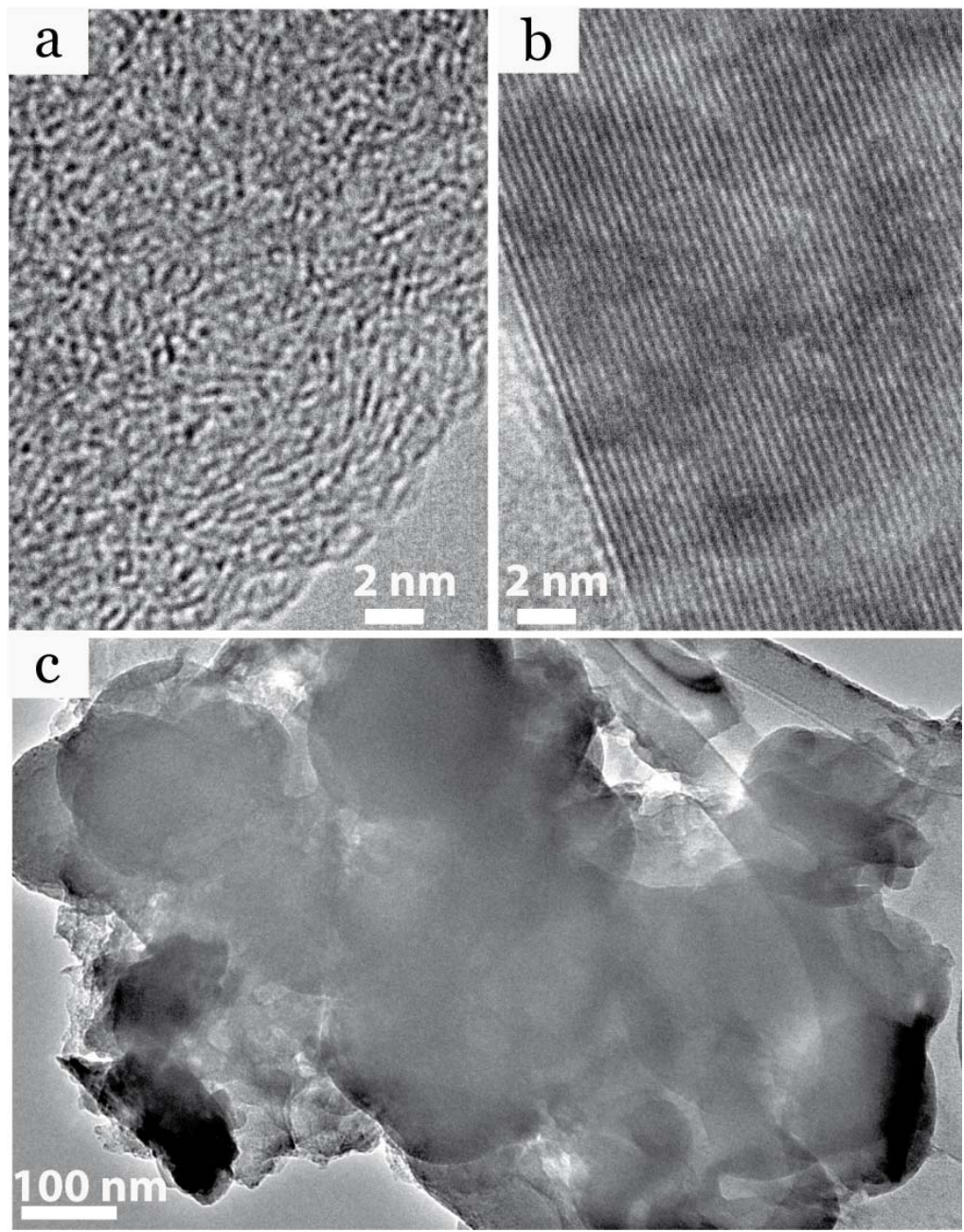

Fig. 12.

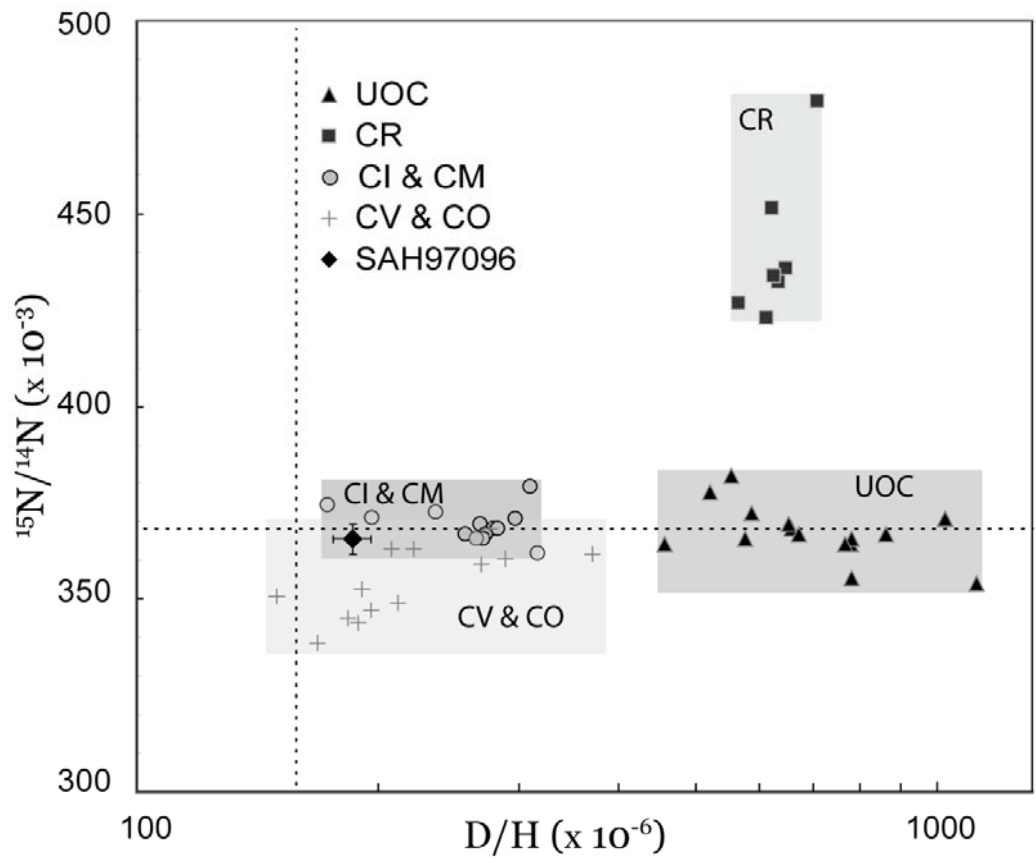

# FIGURES CAPTIONS

Fig.1. Flow-chart of the experimental process with links toward the related method paragraphs.

Fig. 2. (a) The sample was pasted over a support for polishing; (b) it was unstuck and fixed on an aluminium disk with an aluminium ring, (c) silver paint permits to avoid the charging effects.

Fig. 3. SEM picture of the fined grained matrix located around a chondrule (Ch) and composed of silicates (S), kamacite (K), sulphides (Tr: troilite, O: oldhamite, N: niningerite), schreibersite (Sch) and perryite (P).

Fig.4. (a) Optical microscopy image and (b) the corresponding Raman map of a matrix area of Sahara 97096 showing the location of the organic matter (red or blue) and the enstatite (green). Green pixels characterize enstatite by the presence of the main vibrational band of the enstatite at ~1010 cm$^{-1}$, blue pixels the presence of graphitized organic matter with a sharp and intense vibrational band at 1600 cm$^{-1}$ and weak defect bands, and red pixels the presence of disordered organic matter with large vibrational bands having a similar intensity (1350 and 1600 cm$^{-1}$). Graphitized organic matter (blue pixels) is found next to metal or sulfides. (c) Representative Raman spectra of different mineral phases in Sahara 97096. The Raman spectrum of the organic matter inclusion in metallic nodules is compared with an organic matter spectrum obtained in the fine grained matrix: organic matter in metal shows a highly graphitized structure compared to that of matrix organic matter.

Fig.5. SEM images of graphitized carbon inclusions in metal: (a) BSE image of a kamacite (K)-perryite (P)-schreibersite (S) nodule, (b) and (c) two spherulitic graphitized inclusions in metal (the grey squares on (b) and (c) are due to the NanoSIMS imaging).

Fig. 6.
*In situ* images (15 x 15 μm$^2$) of chemical and isotopic ratios calculated from the NanoSIMS analysis of the Sahara 97096 whole rock: (a) carbon and (b) silicon distribution of a matrix area. Sub-micrometer carbonaceous grains are widespread within the silicates of the matrix. (c) nitrogen isotopic distribution and (d) silicon distribution of a graphitized carbon inclusion in a metallic nodule with kamacite (K), perryite (P) and silicon nitride (SN). A Si-rich inclusion (probably silicon nitride) can be observed on the top left corner of the NanoSIMS images.
Images (20 x 20 μm$^2$) of chemical and isotopic ratios calculated from the NanoSIMS analysis of the Sahara 97096 IOM (altered part): (e) C/H ratio, (g) C/N ratio, (f) D/H and (h) $^{15}$N/$^{14}$N ratio. The elemental ratios and isotopic ratios are expressed relative to a terrestrial type III kerogen from Mahakam Delta and used as an internal standard. $^{15}$N-hospots (yellow arrows) and $^{15}$N-coldspot are shown on (h). The black portion on the images corresponds to the non-organic parts that contain less than 15% of the maximum carbon concentration in the image (expressed in counts per second of $^{12}$C$^-$).
Images (20 x 20 μm$^2$) of chemical and isotopic ratios calculated from the NanoSIMS analysis of the Sahara 97096 IOM (preserved part): (i) Nitrogen isotopic distribution and (j) silicon distribution in the IOM of Sahara 97096 obtained by NanoSIMS imaging during the second session of analyses, reveal two types of hotspots: a silicon-rich with a low $^{15}$N enrichment (black arrow) and two silicon-poor with high $^{15}$N enrichment (white arrow). The image is the result of the sum of 100 cycles obtained during the analysis.

Scale bars = 5 μm.

Fig. 7. Distribution of the N/C ratio in the ROI from 6 ion images of the Sahara 97096 IOM (4213 ROI) and from 4 ion images of the graphitized organic matter in metallic nodules (2237 ROI) on frequency histograms with bin sizes of 0.01. Each ROI is ~ 0.35 μm$^2$.

Fig. 8. The 650°C Curie point pyrochromatogram of SAH97096 IOMp. Pyrolysis products are mostly aromatic hydrocarbons. Star indicates a non identified peak. Carboxylic acids could be terrestrial contaminants.

Fig. 9. (a) Room temperature EPR spectra of the Sahara 97096 IOM and (b) expanded view of the field range corresponding to the radicals. Their signal is not detected.

Fig. 10. Raman spectrum of the IOM in the isolated IOM with the two main bands: D band at ~ 1355 cm$^{-1}$ and the G and D2 bands composing the broad band around 1600 cm$^{-1}$.

Fig. 11. HRTEM images of the IOM from SAH97096: (a) disordered organic matter, (b) highly graphitized lamellae and (c) the aggregate of pseudo-amorphous carbonaceous spheres.

Fig.12. The D/H and $^{15}N/^{14}N$ ratios of SAH97096 IOM are compared with those of carbonaceous and unequilibrated ordinary chondrite IOM (from Alexander et al. 2007). The presence of SAH97096 in the CI & CM domain and in the CV & CO domain reinforces the idea that SAH97096 corresponds to a thermal evolution of primitive CI-, CM-like IOM. Dashed lines indicate the terrestrial standard references for D/H and $^{15}N/^{14}N$ ratios (the standard mean ocean values and atmosphere value respectively).